\documentstyle[11pt,epsf]{article}
\newcommand{\ewxy}[2]{\setlength{\epsfxsize}{#2}\epsfbox[10 60 640 570]{#1}}

\let\ga=\gamma

\let\de=\delta
\let\De=\Delta

\let\del=\nabla
\let\si=\sigma

\let\om=\omega
\let\Om=\Omega

\def\to{\rightarrow}
\let\p=\partial
\let\<=\langle
\let\>=\rangle

\let\txt=\textstyle

\let\ad=\dagger

\def\eqn#1{(\ref{#1})}  
\def\e{ {\rm e} }

\def\beq{\begin{equation}}
\def\eeq{\end{equation}}
\def\ba{\begin{array}}
\def\bea{\begin{eqnarray}}
\def\ea{\end{array}}
\def\eea{\end{eqnarray}}

\def\slash{\!\!\!\!/\,}
\def\sl{\slash}

\def\dag{\dagger}

\def\comment#1{ \hbox{[{\it Comment suppressed here.}\/]} }
\def\hide#1{}
\def\o{\over}      
\def\C{ {\cal C} }
\def\O{ {\cal O} }
\def\H{ {\cal H} }
\def\S{ {\cal S} }

\def\Tr{\hbox{Tr}}
\def\P{\hbox{P}}
\def\Ord{ {\rm O} }

\def\Re{ {\rm Re}\, }
\def\x{{\bf x}}
\def\y{{\bf y}}
\def\z{{\bf z}}

\def\IR{\relax{\rm I\kern-.18em R}}
\def\IN{\relax{\rm I\kern-.18em N}}
\def\IB{\relax{\rm I\kern-.18em B}}
\def\IE{\relax{\rm I\kern-.18em E}}
\def\ZZ{\relax{\sf Z\kern-.4em Z}}

\def\TT{\mathchoice
       {\sf T\kern-0.52 em{T}}{\sf T\kern-0.52 em{T}}
       {\sf T\kern-0.40 em{T}}{\sf T\kern-0.40 em{T}}}
\def\IP{\mathchoice
       {\sf I\kern-0.14 em{P}}{\sf I\kern-0.14 em{P}}
       {\sf I\kern-0.11 em{P}}{\sf I\kern-0.11 em{P}}}
\def\id{1\kern-.25em {\rm l}}

\def\ontopss#1#2#3#4{\raise#4ex \hbox{#1}\mkern-#3mu {#2}}

\newcommand{\skipover}[1]{}
\newcommand{\nn}{\nonumber \\}

\def\half  {{\txt {1\over 2}}}
\def\third {{\txt {1\over 3}}}
\def\fourth{{\txt {1\over 4}}}
\def\threefourth{{\txt {3\over 4}}}
\def\sixth{{\txt {1\over 6}}}
\def\fivesixth{{\txt {5\over 6}}}

\def\s={\! = \!}        
\def\sp{\! + \!}  
\def\sm{\! - \!}  
\def\={\,=\,}
\def\+{\,+\,}
\def\-{\,-\,}

\def\bA {\bar{A}}
\def\bU {\bar{U}}

\def\mn {{\mu\nu}}
\def\Fmn{F_\mn}

\def\psib{{\bar\psi}}
\def\rhob{{\bar\rho}}
\def\Omb{{\bar\Om}}

\def\Dssl{{{\bf D}\slash}}
\def\sigF{\si \! \cdot \! F}

\def\BCs{boundary conditions}
\def\SF{Schr\"odinger functional}

\newdimen\pmboffset
\def\oldpmb#1{\setbox0=\hbox{#1}%
 \copy0\kern-\wd0
 \kern\pmboffset\raise 1.732\pmboffset\copy0\kern-\wd0
 \kern\pmboffset\box0}

\def\appendix{\par                              
    \setcounter{section}{0}                     
    \setcounter{subsection}{0}
    \renewcommand{\theequation}{\Alph{section}.\arabic{equation}}
    \renewcommand{\thesection}{Appendix \Alph{section}
                \setcounter{equation}{0}  } 
}

\catcode`\@=11

\def\section{
\setcounter{equation}{0}        
\@startsection {section}{1}{\z@}{-3.5ex plus -1ex minus 
  -.2ex}{2.3ex plus .2ex}{\Large\bf}}
\renewcommand{\theequation}{\arabic{section}.\arabic{equation}}

\def\subsection{\@startsection{subsection}{2}{\z@}{-3.25ex plus -1ex minus 
 -.2ex}{1.5ex plus .2ex}{\normalsize\bf}}

\def\subsubsection{\@startsection{subsubsection}{3}{\z@}{-3.25ex plus
 -1ex minus -.2ex}{1.5ex plus .2ex}{\normalsize}}

\def\@eqnnum{%
\savebox{\eqnumb}{\rm (\theequation)}%
\settowidth{\numblen}{\usebox{\eqnumb}}%
\makebox[\numblen][l]{\usebox{\eqnumb}~~~\usebox{\eqlabel}}}

\catcode`\@=12

\newsavebox{\eqlabel}

\catcode`\@=11
\newlength{\numblen}
\newsavebox{\eqnumb}
\def\@eqnnum{%
\savebox{\eqnumb}{\rm (\theequation)}%
\settowidth{\numblen}{\usebox{\eqnumb}}%
\makebox[\numblen][l]{\usebox{\eqnumb}~~~\usebox{\eqlabel}}%
}
\catcode`\@=12

\newenvironment{equationwithlabel}[1]{ %
  \savebox{\eqlabel}{#1}
  \begin{equation}\label{#1} }{\end{equation}\savebox{\eqlabel}{~}}

\newcommand{\beql}[1]{\begin{equationwithlabel}{#1}}
\newcommand{\eeql}{\end{equationwithlabel}}

\newenvironment{eqnarraywithlabel}[1]{ %
  \savebox{\eqlabel}{#1}
  \begin{eqnarray}\label{#1} }{\end{eqnarray}\savebox{\eqlabel}{~}}

\newcommand{\beal}[1]{\begin{eqnarraywithlabel}{#1}}
\newcommand{\eeal}{\end{eqnarraywithlabel}}
\begin{document}

\begin{flushright}
FSU-SCRI-97-43 \\
May 1997 \\
\end{flushright}
\vskip 6mm
\begin{center}
{\Large \bf The Schr\"odinger Functional for\\[1.1mm] 
Improved Gluon and Quark Actions}
\vskip 7mm
{\normalsize Timothy R. Klassen}
\vskip 0.5mm
{\normalsize  SCRI, Florida State University\\[0.7mm]
     Tallahassee, FL 32306-4052, USA}
\vskip 6mm

{\normalsize \bf Abstract}

\vskip 4mm

\begin{minipage}{5.0in}  
{\small 
The \SF{} (quantum/lattice field theory with Dirichlet boundary 
conditions) is a powerful tool in the non-perturbative 
improvement and for the study of other aspects of lattice QCD.
Here we adapt it      to improved gluon and quark
actions, on isotropic as well as anisotropic lattices.
 Specifically, we describe the structure of the boundary layers,
obtain the exact form of the classically improved gauge action,
and outline the modifications necessary on the quantum level.
The projector structure
of Wilson-type quark actions determines which field components can
be specified at the boundaries. We derive the form of $\Ord(a)$
improved quark actions and describe how the coefficients can be
tuned non-perturbatively. There is one coefficient to be tuned
for an isotropic lattice, three in the anisotropic case.\newline
Our ultimate    aim is the construction of actions 
that allow accurate simulations of all aspects of
QCD on coarse lattices.
}

\end{minipage}
\end{center}
\vskip 6mm

\renewcommand{\thepage}{\arabic{page}}
\setcounter{page}{1}


\section{Introduction}

It has become generally recognized in the last few years that accurate
continuum extrapolations of lattice QCD will be possible in the
foreseeable future only by using improved actions (see e.g.~\cite{LAT96proc}).

In the Symanzik (on-shell) improvement program~\cite{Sym,LWGlue,SW,May,ILQA},
which we will follow, one systematically adds higher dimensional operators
to an action, to cancel its lattice artifacts to some order in the
lattice spacing $a$. Composite operators can be similarly improved.
The coefficients of the improvement terms  are         trivial to
calculate at tree-level; one-loop calculations are also possible,
but already much more difficult, in general. Even if known, perturbative
improvement coefficients per se are of limited value, since
naive lattice perturbation theory does not work very well. It dramatically 
underestimates the improvement coefficients, for example.

 Tadpole improved perturbation theory~\cite{LM} works much better. For
gluonic actions, where the leading errors of the Wilson plaquette
action are $\Ord(a^2)$, tree-level (or one-loop) plus tadpole improvement
seems to cancel most of these errors, allowing quite
accurate calculations on coarse lattices ($a=0.1-0.4$~fm, say), as several
studies~\cite{Alf1,BiTD,Bliss,Ukawa} have 
demonstrated.
 Similarly, it seems that the leading $\Ord(a^2)$ errors of the Wilson or 
Sheikholeslami-Wohlert (SW)~\cite{SW} quark
actions, which break rotational symmetry, can be cancelled to a large 
extent with classical and tadpole improvement. This is demonstrated
by the excellent dispersion relations~\cite{LAT95,LAT96} of the 
D234~\cite{ILQA} actions, for example.

On the other hand, for Wilson-type quark actions there are, at least
at the quantum level, also $\Ord(a)$ violations of chiral symmetry.
Even without the explicit non-perturbative determination of the $\Ord(a)$
improvement coefficient of the SW action (see below), there was clear 
evidence~\cite{Bi96} that this coefficient is not estimated sufficiently 
accurately by tadpole improvement on coarse lattices. 
In any case, it is obviously
preferable to determine at least the leading improvement coefficients
non-perturbatively.

In seminal work, described in a series of papers (of which some 
are~\cite{Dec95,May,Sep}), 
the ALPHA Collaboration has 
recently demonstrated how to do this.
Using the \SF~\cite{SymSF,LNWW}  and the demand that the PCAC relation
hold for small quark masses, they 
determined the $\Ord(a)$ improvement coefficient of the SW quark action 
(as well as other quantities) on quenched Wilson glue.
They find that it is almost 20\% larger than the (plaquette) tree-level
tadpole estimate even for a relatively fine $a\s= 0.1$~fm lattice.

Motivated by these observations, our aim in this paper is to adapt the
\SF{} to improved gluon and quark actions. 

The term
``\SF'' is another name for the partition function of a euclidean quantum
field theory with Dirichlet boundary conditions. In other words, one considers
a theory with fixed \BCs{} on the fields at times $x_0=0$ and $x_0=T$, say.
The \SF{} is a powerful tool not just for the calculation
of $\Ord(a)$ improvement coefficients of Wilson-type quark actions, but
also for the determination of
normalization and improvement coefficients of currents,
as well as running couplings and quark masses. 
This is accomplished,
in short, by using the finite volume and the boundary values of the gauge
and quark fields as probes of the theory. Since Ward identities 
are {\it local} equations, they must hold       independent
of volume and \BCs. Probing the theory in this manner provides constraints
that can be used to determine improvement coefficients.
For the determination of running couplings one uses finite-size scaling
techniques.   One of the more 
technical reasons for the success of the \SF{} is that the use
of non-periodic \BCs{} in the time direction implies the absence of
zero modes (at least for sufficiently small lattice spacings), 
allowing one to perform simulations directly at zero (or small)
quark mass. This avoids a number of conceptual and practical complications.

For the case of pure gauge theory in the continuum it is quite clear what is 
meant (at least in a formal sense) by Dirichlet boundary conditions,
\beq
 {\cal Z}[C',C] \= \int \prod_{\x} \Lambda(\x) \int D[A] \, 
          \e^{-S_g[A]} \, , \qquad
     S_g[A] \= {1\o 4 g^2} \int d^4 x \, \Fmn^a(x) \Fmn^a(x) \, .
\eeq
Here the path integral is over all gauge fields $A_\mu(x)$, $0\leq x_0 \leq T$,
with boundary values ($k=1,2,3$)
\beq
 A_k(x) ~=~ \left\{ 
\begin{array}{ll}
     C_k^\Lambda(\x)         & \quad \mbox{at $x_0=0$} \nn[1mm]
     C'_k(\x)        & \quad \mbox{at $x_0=T$} \, ,
\end{array}  \right.
\eeq
where, generally, $C^\Lambda$ denotes the gauge field obtained after 
applying a gauge transformation $\Lambda$ to $C$.\footnote{The integral over
gauge transformations $\Lambda(\x)$ at $x_0\s= 0$ (or $x_0\s= T$)
is necessary to ensure the gauge invariance of the \SF{}.}

For lattice QCD it is not immediately obvious how Dirichlet \BCs{} should
be defined. The \SF{} for pure gauge theory, in the form of the Wilson
plaquette action, was studied in~\cite{LNWW}; Wilson quarks
were treated in~\cite{Sint}. For the case of improved gluon and quark
actions considered here, we will see that the ``boundary'' in the \SF{}
will consist of a double layer of time slices. This causes slight 
complications at intermediate stages, but ultimately 
the \SF{} formalism for improved actions 
proves to be hardly more complicated than for
Wilson gluons and quarks.

There are several issues concerning the aims, scope and motivation of this 
paper that we should clarify at this point.      First of all, the
phrase ``improved action'' in the above is meant to refer to actions
where at least the classical $\Ord(a^2)$ errors have been eliminated by the
addition of third and/or fourth order derivatives.
Of course, in general it is inconsistent to consider improvement at
order $a^2$ before having eliminated all errors at $\Ord(a)$. However, it
is perfectly consistent to consider classical improvement to {\it any}
order and then supplement it with quantum improvement to some lower
order. It also makes sense to supplement classical with tadpole 
improvement.\footnote{Note, though, that for dynamical QCD,
where the quark dynamics feeds back into the glue, it might be 
necessary to use an iterative procedure to self-consistently tune
the tadpole factors and the $\Ord(a)$ coefficient(s) of the quark action.
It seems likely that this iteration will have converged to sufficient 
accuracy after just a few steps.}

For pure gauge theory perturbative improvement coefficients 
have been calculated to one loop for two actions~\cite{LWGlue,Sym2},
so far. It is conceivable that at some point a practical method will be found
to tune {\it all} gluonic $a^2$ errors to zero (at least for isotropic
lattices).    For fermions, on the other
hand, this is clearly impossible; there are just too many terms
at order $a^2$~\cite{SW}. 
This leads to an obvious question: If we can only eliminate
$\Ord(a)$ but not all $\Ord(a^2)$ errors, why should we even bother to 
eliminate
what we hope are the leading $a^2$ errors via classical and tadpole 
improvement? We will argue that there are both theoretical and
``empirical'' reasons to suggest that this is a worthwhile enterprise.
Let us discuss these reasons in turn. [The reader not interested
in motivational issues, but only in the \SF{} per se, 
can at this point skip ahead to the outline at the end of this section.]

Remember the special nature of the $\Ord(a)$ errors of Wilson-type quarks.
They arise because we added a second-order derivative term to solve the doubler
problem. This term breaks chiral symmetry, and, at least at the quantum
level, $\Ord(a)$ errors become unavoidable.\footnote{If other terms are 
suitably added by a field transformation the chiral symmetry breaking can 
be avoided at the {\it classical} level to basically arbitrarily high order;
cf.~\cite{ILQA} and appendix~C.}
On the other hand, these terms do not break rotational symmetry at $\Ord(a)$
or $\Ord(a^2)$. This is in contrast to the leading $a^2$ errors of both
gluon and quark actions, which break rotational but not chiral symmetry.
One might therefore argue that it is somewhat besides the point to label
these errors as $\Ord(a)$ and $\Ord(a^2)$ errors. Viewing them instead as the 
leading violations of chiral, respectively, rotational symmetry, it makes
perfect sense to try to eliminate both of them as best as one can.

Now to the empirical reasons. Here the basic point is simply
that the elimination of the leading $\Ord(a^2)$ errors through classical and 
tadpole improvement works quite well. We already referred to the surprisingly
accurate results obtained with improved gluon actions on coarse lattices.
Concerning the more difficult case of $\Ord(a^2)$ improved quark actions,
I argued in~\cite{Bi96}, after surveying the available data, that all
quantities with a weak dependence on the $\Ord(a)$ improvement term
(dispersion relations, mass ratios)
are also accurate to a couple percent on coarse lattices when calculated
with such an action. It is only quantities with a relatively strong
dependence on this term (like the $\rho$ mass in units of a gluonic scale
such as the string tension, or the hyper-fine splitting of heavy quark systems)
that still have significantly larger errors on the coarsest lattices.
It is therefore not unreasonable to believe that with no $\Ord(a)$ and
only small $\Ord(a^2)$ errors accurate simulations of QCD will be possible
on coarse lattices. 

Concerning the gluon actions there is another more technical, but potentially
important, issue we should mention. 
Namely, in addition to the $\Ord(a^2)$ errors
of the Wilson plaquette action, there also seem to be
large non-power-like lattice artifacts that set in on lattices with
spacings slightly above  $0.1$~fm. 
What we are referring to is the non-trivial
phase structure of the plaquette action when embedded in a 
larger space of couplings (fundamental and adjoint representations of the
plaquette). This was the subject of many studies in the early days of
lattice gauge theory; cf.~\cite{Urs} for recent results and references.
The (in)famous dip of the discrete $\beta$-function, a similar dip in the
$0^{++}$ glueball mass, and non-smooth behavior in various thermodynamic
quantities are all believed to be consequences of a second-order phase
transition close to the fundamental coupling axis of the Wilson gauge
action. Although further studies are needed, there is 
evidence  that these lattice artifacts are significantly smaller 
with improved 
gluon actions~(see e.g.~\cite{Ukawa,CMnew}). There are also
more direct indications that the second-order phase transition point
moves away from the fundamental coupling axis with improvement~\cite{Steph}.
Finally, it seems that the variance of the static potential 
measured with improved glue~\cite{Alf1,QCDTARO} is much smaller
than for Wilson glue at the same lattice spacing. This might
also be related to the phase structure (on the other hand, there
might be more prosaic reasons for this; this remains to be investigated).

The potential significance of these issues is as follows.
In their study of the $\Ord(a)$ improvement of the SW action
on quenched Wilson glue L\"uscher et~al found~\cite{Sep} that the improvement
coefficient could not be determined for lattices coarser than $0.1$~fm,
due to large fluctuations (``exceptional configurations'') of the glue,
leading to accidental zero modes at small quark mass. In the quenched
approximation this problem is bound to occur at some sufficiently coarse
lattice spacing, but it is not clear  a priori exactly where
it should occur. It could depend on the specific lattice action used.
The   observations described in the previous paragraph, 
lead one   to hope that with improved glue\footnote{And perhaps 
order $a^2$ improved quark actions, which in our experience have less problems 
with ``exceptional configurations''. (Which, by the way, means that
the latter term is strictly speaking a misnomer; the occurrence of near
zero modes depends also on the specific quark action used for a given
configuration.) }
 the breakdown will          occur on coarser lattices.

While finishing the write-up of this paper,
hard facts began to emerge that support the above arguments:
As will be described in~\cite{EHK},
the breakdown of the quenched approximation does
indeed occur on coarser lattices when improved glue is used.

The hope that this would happen was the main motivation underlying this
paper. To obtain accurate continuum results from lattice QCD requires
simulations at a series of lattice spacings that provide a significant
``lever arm'' for extrapolations. Due to the dramatic increase in the
cost of a QCD simulation as the lattice spacing is decreased, such
a series can be obtained orders of magnitude cheaper if one makes
use of coarse lattices.

Concerning the scope of this paper we should mention that we will also
consider {\it anisotropic} lattices~\cite{ILQA,Kar,aniso}. 
Such lattices, with a smaller temporal than spatial lattice spacing, 
offer the best hope to tackle certain hard problems, like the simulation
of heavy quarks in a relativistic formalism~\cite{LAT96,ILQA} and detailed
studies of glueballs~\cite{CMnew}.
Due to the breaking of manifest space-time exchange
symmetry on anisotropic lattices, there are more terms that have
to be tuned to eliminate all $\Ord(a)$ errors of a Wilson-type quark action
(as we will discuss). Tuning these errors to zero is therefore more 
complicated, but should be within reach of the \SF{} technology
(cf.~sect.~\ref{sec:PCAC}).  Note that
anisotropic lattices fit quite naturally into the \SF{} framework, since the
boundary conditions already treat space and time differently.

The outline of this paper is as follows. In sect.~2 we derive the \SF{}
for improved gluon actions. We discuss the structure of the boundary 
layers and the precise form of the (classical) action at the boundaries, 
where the fields are given fixed values. 
We then specialize to spatially constant boundary values and 
discuss the classical background field they induce in the bulk, including 
its uniqueness.
The subject of sect.~3 is Dirichlet \BCs{} for (improved) quark actions.
All Wilson-type quark actions of practical interest, on
isotropic as well as anisotropic lattices, have a certain 
projector structure, which dictates which field components can
be fixed at the boundaries.
Once these are specified, one can define boundary quark fields as 
functional derivatives with respect to the boundary values.
These boundary fields are extremely useful in applications of the 
\SF. One application is described in sect.~4, namely, the determination
of the $\Ord(a)$ improvement coefficients of Wilson-type quark actions.
Concluding remarks and a brief discussion of future work
can be found in sect.~5.

Appendix~A summarizes some of our notation and conventions. In appendix~B
we discuss the classical boundary errors of the pure gauge \SF.
In appendix~C, finally,
we derive the form of $\Ord(a)$ on-shell improved quark actions on 
anisotropic lattices. We also recall~\cite{SW,ILQA} the classical 
improvement  of quark actions.

This paper is largely self-contained, but the reader might want to 
consult~\cite{LNWW} for a detailed introduction to the \SF,
ref.~\cite{Dec95} for a summary of various applications, 
and~\cite{May,ILQA} for discussions of 
Symanzik improvement for fermions.

\section{Gauge Fields}\label{sec:gluons}

\subsection{Gauge Actions in the Bulk}

Consider a four-dimensional hypercubic lattice of extent 
$L_\mu$ 
and lattice spacing $a_\mu$ in direction $\mu=0,1,2,3$.
Except for purely classical considerations
we assume the lattice to be spatially isotropic
with $a\equiv a_k, \, k\s= 1,2,3$.
Other notations and conventions are summarized in appendix~A. 

To avoid having to worry about boundary effects at this point, let us
impose periodic \BCs{} on the gauge fields for the moment. The lattice
gauge actions of interest can be written as
\beq\label{gauge_actions}
S_g[U] \= {2\o g^2} \sum_\tau \sum_{\C\in\S_\tau}  
                        c_\tau \,  \Re\Tr\,[1\- U(\C)] \, ,
\eeq
where $U(\C)$ denotes the product of link fields along the loop $\C$,
$\tau$ labels a finite set of types of loops, and $\S_\tau$ is the
set of all positively oriented (say) loops of a given type that can
be drawn on the lattice. On an isotropic lattice the types of loops
might include plaquettes, rectangles, etc; on an anisotropic lattice
one would have to distinguish spatial plaquettes, temporal plaquettes,
spatial rectangles, short-temporal rectangles, long-temporal rectangles, etc.

The real numbers $c_\tau$ are determined by some improvement condition. We do
not really need to
know the exact numbers to set up the \SF, but for concreteness let us
be more explicit about a simple class of actions that is sufficient to 
illustrate what happens for more general improved actions. Namely, for
``plaquette $+$ rectangle'' actions~\cite{PR} we can write
\beq\label{PRactions}
 S_g[U] \=   
{2\o g^2}\, \sum_{x,\,\mu < \nu}
{\Delta V\o a_\mu^2 a_\nu^2} \,
\Re\Tr\, \left [
   c^{(0)}_{\mn} \, \frac{1-P_{\mn}(x)}{u_{\mu}^2u_{\nu}^2}
\- c^{(1)}_{\mu} \, \frac{1-R_{\mn}(x)}{u_{\mu}^4u_{\nu}^2}
\- c^{(1)}_{\nu} \, \frac{1-R_{\nu\mu}(x)}{u_{\nu}^4u_{\mu}^2}
\right ] \, ,
\eeq
where $\Delta V = \prod_\mu a_\mu$ is the volume of a single lattice hypercube,
and $P_\mn$ and $R_\mn$ are equal to $U(\C)$ 
with $\C$  a $1\times1$ (plaquette), respectively, $2\times1$ (rectangle) 
Wilson loop in the $\mn$ plane. 
The coefficients $ c^{(0)}_{\mn} = 1 + 4(c^{(1)}_{\mu} +  c^{(1)}_{\nu})$ and
for good measure we have also indicated tadpole improvement factors 
$u_\mu$ that one might want to include.\footnote{On anisotropic lattices 
additional renormalization factors might have to be included, depending
on how one defines the tadpole factors.}
As special cases this class of actions contains the Wilson and tree-level
$\Ord(a^2)$ improved~\cite{PR,LWGlue} actions,
\beq\label{c_one}
 c^{(1)}_{\mu} ~=~ \left\{ 
\begin{array}{cl}
      0                       & \quad \mbox{Wilson action} \nn[1mm]
   {1\o 12}                   & \quad \mbox{Improved action} \, .
\end{array}  \right.
\eeq              
Another interesting case, on anisotropic lattices, is defined by
$c^{(1)}_{\mu}={1\o 12} (1-\delta_{\mu 0})$ (cf.~\cite{aniso}). 
Only the spatial directions are improved now; there are no
tall-temporal rectangles in the action, leading to $a_0^2$ errors. 
Since this action contains only loops of maximal temporal extent $1$,
there are      no unphysical branches in its dispersion relation.


\subsection{Transfer Matrix}

The \SF{} can either be defined as the partition function with fixed
boundary conditions, or as a suitable power of the transfer matrix evaluated
between initial and final states. On the lattice the meaning of ``fixed
boundary conditions'' is not a priori clear.  Understanding this issue
via the transfer matrix is at least pedagogically useful, so we will
proceed by reviewing the transfer matrix formalism. For improved gauge
actions this formalism was described in~\cite{LWTM}, whose approach we 
will follow.

We will use the plaquette $+$ rectangle actions as examples, but as we
proceed it will become clear how to construct the transfer matrix for
more general actions.
We will also see that the general form of the transfer 
matrix is determined solely by the maximal temporal extent of the loops 
appearing in the action. 
To avoid tedious and irrelevant case distinctions, we will 
use the term ``Wilson action'' as a shorthand for any gauge
action with loops of maximal temporal extent $1$.
Note that this also includes plaquette $+$ rectangle actions without
tall-temporal rectangles, which, as mentioned earlier,
are useful for anisotropic lattices.   We will use
the phrase ``improved action'' to denote actions with loops of maximal 
temporal extent $2$.

For pedagogic reasons we       first consider the Wilson case.
 The classical lattice field equations for such actions
(in temporal gauge $U_0(x)\equiv 1$, say) are second order difference equations
in $x_0$. Specifying the link field on one initial and one final time slice 
leads to a unique solution to the field equations. The
transfer matrix of the quantum theory should therefore give
the amplitude to go from some gauge field on one 
time slice to another field on some other time slice.

To  define the transfer matrix we first have to specify the Hilbert
space ${\cal H}$ on which it is supposed to act. We take $\H$ to consist
of wave functions $\varphi[U]$ that are functions of the gauge field $U$
on a fixed time slice, with the natural inner product
\beq
 \< \varphi_2 | \varphi_1 \> \= \int D[U] ~ \varphi_2[U]^\ast \,\varphi_1[U] ~,
    \qquad\quad D[U] \, \equiv \, \prod_{{\bf x}, k} \, dU_k({\bf x}) ~,
\eeq
in terms of the Haar measure $dU$ on SU($N$).
More precisely, $\H$ consists of all gauge invariant wave functions of
finite norm.
Note that any wave function can be made gauge invariant by averaging over
gauge transformations $U\to U^\Lambda$. This defines a projection operator
onto the physical subspace,
\beq
  \IP \varphi[U] \= \int \prod_{\x} \, d\Lambda(\x) \, \varphi[U^\Lambda] \, .
\eeq
 We can write $\varphi[U] = \< U|\varphi\>$ where the states
$|U\>$ are eigenstates of the link field operators $\hat{U}_k(x)$ in the
Schr\"odinger representation, with eigenvalues $U_k(x)$. 
For gauge invariance the projector $\IP$ should be applied to the
$|U\>$'s, which will be understood in the following.
The identity can be decomposed as $\id = \int D[U] \, |U\>\<U|$.

\begin{figure}[tb]
\vskip -110mm
\mbox{ \hspace{ 1 em}\ewxy{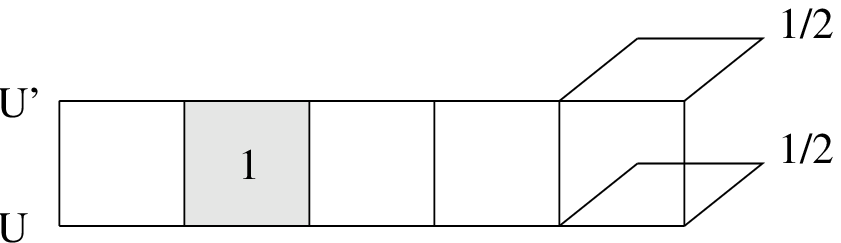}{160mm} 
       \hspace{-21em}\ewxy{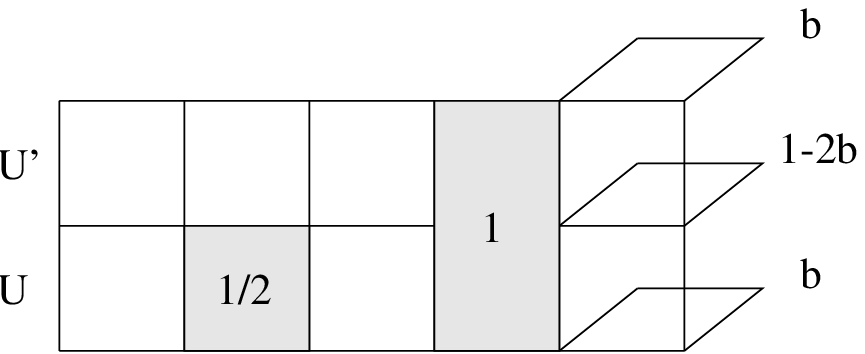}{160mm} } 
\vskip 16mm
\caption{
Temporal weight factors of various loops for Wilson (left)
and improved (right) action transfer matrices.
}
\label{fig:tm}
\vskip 5mm
\end{figure}

The transfer matrix $\TT$ acts on wave functions as
\beq\label{Tphi}
 (\TT \varphi)[U'] \= \int D[U] ~ \<U'|\TT|U\> \, \varphi[U] ~,
\eeq
where its matrix element (or kernel) to go from gauge field $U$ at time $t$
to $U'$ at time $t+a_0$ is given by
\beq
 \<U'|\TT|U\> \= \int \prod_{{\bf x}} dV_0(t,{\bf x}) ~
    \e^{-\Delta S[U',V_0, U]}  ~, \qquad\quad V_0(t,{\bf x})\in{\rm SU}(N) ~.
\eeq
Here $\Delta S[U',V_0, U]$ is given by exactly the same expression as the
action itself, restricted to a two time slice lattice with gauge fields
$U$ at $t$, $U'$ at $t+a_0$, and $V_0$ playing the role of the temporal link
fields\footnote{Note that the integral over $V_0$ will project $\varphi$ 
in~\eqn{Tphi} onto its gauge invariant part, in case it was not gauge invariant
to start with.}  
--- {\it except} that the contribution from loops completely within the 
spatial planes at $t$ and $t+a_0$ are weighted with a factor of $\half$.
This is illustrated in figure~\ref{fig:tm}.
It is clear why this factor of $\half$ must be present: When we take powers
of the transfer matrix (to get the partition function or \SF, see below)
all gauge fields on the internal time slices contribute twice.
Similarly, a correlation function involving gauge fields is equal
to some power of the transfer matrix with operators $\hat{U}$
inserted at appropriate places in the product.

We now turn to the $\Ord(a^2)$ improved case. With loops in the action
extending up to  two steps in the temporal direction, the action involves
second order,
the field equations fourth order temporal differences (in temporal gauge).
There will be a unique solution to the field equations if we specify
the gauge field on two initial and two final time slices. Correspondingly,
we expect the Hilbert space to consist of wave functions on two
consecutive time slices, and the transfer matrix to give the amplitude for 
transitions between such double layers.

Let 
\beq
  \varphi[U] ~\equiv~ \varphi[U^{(+)},U^{(0)},U^{(-)}]
\eeq
be a wave function depending on the link field $U^{(-)}$ on time slice 
$t$, $U^{(+)}$ on time slice
$t+a_0$, as well as the temporal links $U^{(0)}$ connecting the two time
slices. Let the Hilbert space consist of gauge invariant wave functions of 
finite norm, with respect to the inner product
\beq
 \< \varphi_2 | \varphi_1 \> \= \int D[U] ~ \varphi_2[U]^\ast\, \varphi_1[U] ~,
    \qquad D[U] \, \equiv \, \prod_{{\bf x}} \Bigl [ dU^{(0)}({\bf x}) 
 \prod_k \, dU^{(+)}_k({\bf x}) \,  dU^{(-)}_k({\bf x}) \Bigr] \, .
\eeq
The transfer matrix is then defined by
\beq
 \<U'|\TT|U\> \= \e^{-\Delta S[U', U]} ~ \prod_{{\bf x},k} ~
\delta\Bigl( U'^{(-)}_k({\bf x}),\, U^{(+)}_k({\bf x})\Bigr) ~.
\eeq
The $\delta$-function        
 (with respect to the Haar measure)
means that the transfer matrix is effectively a
``$2\to 1$'' and not a ``$2\to 2$'' time slice operator.
$\Delta S[U', U]$ is the action of a three time slice lattice obtained
by fusing the top layer $U^{(+)}$ of $U$ with the bottom layer 
$U'^{(-)}$ of $U'$, where we weigh a Wilson loop of temporal 
extent $\Delta t$ (in lattice units) by a factor 
\beq\label{wgt}
 w(\Delta t\s= 2) = 1 \, , \quad w(\Delta t\s= 1) = {1\o 2} \, , \quad
 w(\Delta t\s= 0) =  \left\{ 
\begin{array}{cl}
      b              & \quad \mbox{at $x_0 = t$} \nn
     1-2b            & \quad \mbox{at $x_0 = t+a_0$} \nn
      b              & \quad \mbox{at $x_0 = t+2a_0$}
\end{array}  \right.
\eeq
This is presented graphically in figure~\ref{fig:tm}.
Note that $\<U'|\TT|U\>$ is gauge invariant, since $\Delta S$ is.

Time reflection symmetry dictates the above form of the 
``temporal weight factors'' $w(\Delta t)$.
When taking powers of the transfer matrix the $w(\Delta t)$ combine
to give $1$ for all loops in the bulk, ensuring that we obtain the correct
(bulk) partition and correlation functions. This is true for any value of
the (real) parameter $b$. The effect of $b$ is only noticeable at the
boundaries. Its value is fixed, classically, by demanding the absence
of order $a^2$ boundary errors.\footnote{In~\cite{LWTM} periodic \BCs{}
were considered, where the issue of boundary errors does not arise, and
the ``symmetric'' choice $b=1/3$ was used. As we will see, this is not
the value minimizing boundary errors with Dirichlet \BCs.}
This will be discussed in the next subsection and appendix~B. Let us
for the moment keep $b$ a free parameter.

The transfer matrix of improved actions is not hermitean~\cite{LWTM}, since
energy eigenvalues are complex at the order of the cutoff (this can also
be seen by studying the dispersion relation). In most situations
the resulting effects, chiefly oscillatory behavior of correlation functions
at small times, do not create significant problems in practice.\footnote{The 
one obvious exception being glueballs, whose bad signal to noise properties
force one to work with the first few time slices. As mentioned in the
introduction, anisotropic lattices can help here.}

\subsection{Gauge Actions with Dirichlet Boundary Conditions}

\begin{figure}[tb]
\vskip -80mm
\mbox{ \hspace{ 1 em}\ewxy{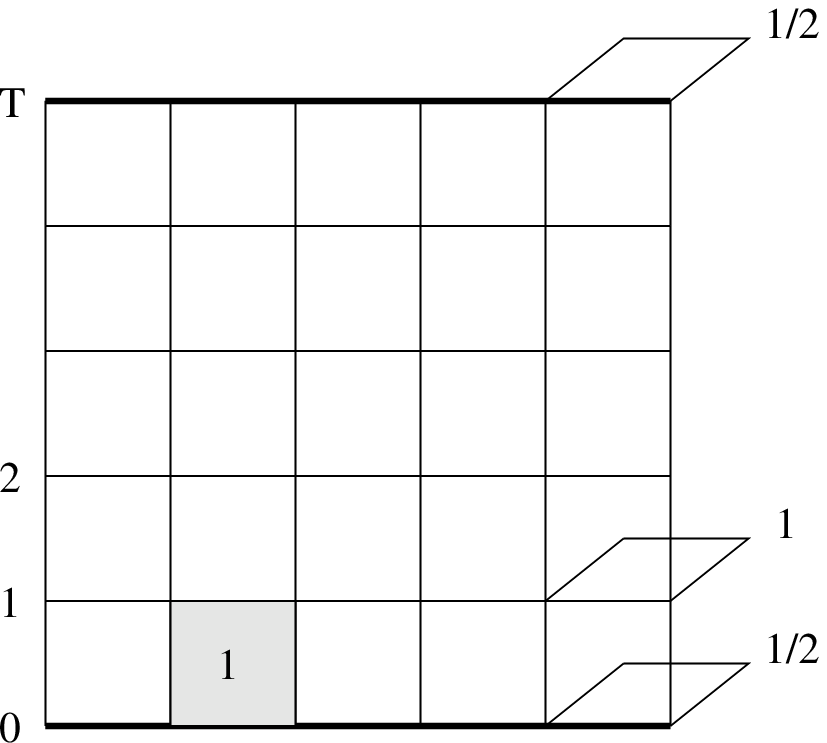}{160mm} 
       \hspace{-21em}\ewxy{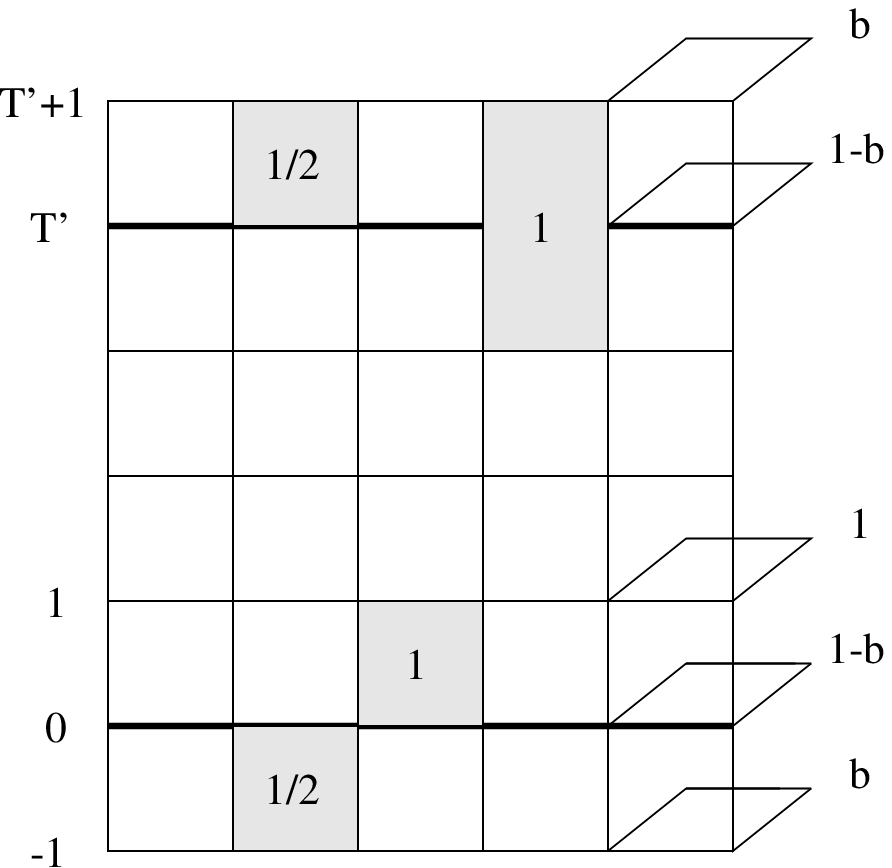}{160mm} } 
\vskip 16mm
\caption{
Temporal weight factors of various loops in the Wilson (left)
and improved (right) actions with Dirichlet boundary conditions on lattices 
with $L=5 a$ and $T=5 a_0$. Time, measured in lattice units, runs upwards, 
as marked on the left ($T'\equiv T-1$ for the improved case).
The thick lines indicate the boundaries of the bulk region;
every loop   with at least one link strictly inside the bulk
has a temporal weight factor of $1$.
}
\label{fig:wgts}
\vskip 8mm
\end{figure}

We can now write down gauge actions with Dirichlet boundary 
conditions. 
We choose the spatial \BCs{} on the gauge fields to be periodic.
To obtain the Schr\"odinger functional on a lattice of
temporal extent $T/a_0$ we just have to evaluate $\TT^{T/a_0}$ between
fixed initial and final states. 
 In the Wilson case the gauge fields have
to be specified at $x_0= 0$ and $x_0=T$, in the improved case on the
$x_0=-a_0,0$ and $x_0=T',T'+a_0$ double layers. To allow for a unified
notation for the Wilson and improved cases in various formulas, we have 
introduced
\beq
       T' ~\equiv ~ \left\{ 
\begin{array}{ll}
      T              & \quad \mbox{Wilson action} \nn[1mm]
     T \- a_0        & \quad \mbox{Improved action} ~.
\end{array}  \right.
\eeq
 Denoting the boundary gauge
fields by $W$ and $W'$, we define the Schr\"odinger functional for
pure gauge theory as
\beq
 {\cal Z}[W',W] \= \< W' | \TT^{T/a_0} | W \> \, .
\eeq
Remember that in the improved case $W$ is a shorthand for the fields
$W^{(+)},W^{(0)},W^{(-)}$ on a double layer (and ditto for $W'$).
More explicitly, we can write
\bea
{\cal Z}[W',W] &\s=& \int D[U] \, e^{-S_g[U;W',W]} \,\, , \qquad
 D[U] \,\equiv \, \prod_{\x} \, \Biggl[ dU_0(0,\x) \, 
    \prod_{x_0=a_0}^{T'-a_0} \, \prod_\mu \, dU_\mu(x) \Biggr ]\nn[.6mm]
S_g[U;W',W] &\s=& {2\o g^2} \sum_\tau \sum_{\C\in\S_\tau} \, w(\C)\, c_\tau \, 
   \Re\Tr\,[1\- U(\C)] \, ,
\eea
where the temporal weight factor $w(\C)$
for a loop in an action with Dirichlet
\BCs{} is best defined pictorially, as in figure~\ref{fig:wgts}.

As remarked in the caption of this figure, all loops containing a
dynamical link have a temporal weight factor of~$1$. This fact simplifies
the coding of the updating algorithm for the \SF.
 
We now discuss the boundary errors of the \SF. In particular we have to
fix the so far free parameter $b$ in the improved case. 
In view of the remark in the previous
paragraph, the reader might wonder at this point, how errors associated 
with the boundary fields, which are, after all, fixed in a
simulation, can have any effect on an observable defined in terms
of dynamical variables in the bulk. 

First of all, even on the classical level 
it is legitimate to consider observables defined by taking functional
derivatives with respect to the boundary values, and only then fixing
them at some suitable value. 
This introduces cross terms between the bulk and the boundary fields.
The ``SF coupling'' defined in~\cite{gSF},
for example, does involve such derivatives.
Secondly, on the quantum level the boundary effects can ``spill into
the bulk'' and affect the dynamics there, as we will see shortly.

Let us first consider the  classical boundary errors
of the \SF. We  want to know the difference between the 
lattice and the continuum actions with corresponding \BCs.
This difference has two sources. One is the difference between the
continuum and lattice action {\it densities} (the integrand/summand), the 
other the difference between an integral and a discrete sum approximating
it. 

This is basically a classical calculus problem; the
elementary, if slightly wordy, tracking of the differences for the
Wilson and improved actions is discussed in appendix~B. We here only
state the conclusion: The Wilson action has order $a^2$ boundary
errors, and the improved action only $a^4$ errors if we choose $b=1/24$.

Actually, at tree level the precise value of $b$ does not matter
for the special (spatially constant) boundary values of the gauge field
we will consider later. For many applications
this is even true on the quantum level at $\Ord(a)$; 
a point to which we will return in a moment.

In the quantum case it is expected~\cite{SymSF,LNWW} 
that in addition to the usual
counter terms in the bulk, some small number of additional counter terms
might be necessary on the boundaries. In pure  gauge theory the only possible
terms at $\Ord(a)$ are spatial sums over the boundary time slices
of lattice versions of
$\Tr\, F_{0k}^2$ and  $\Tr\, F_{kl}^2$. Note that there is no
difference in this respect between isotropic and anisotropic lattices
(in contrast to the bulk, where fewer counter terms are necessary in the
isotropic case).

So, no terms not already present in the classical action
are required on the quantum level. 
In addition to the usual renormalization of the gauge coupling in the bulk
(and other bulk renormalizations for anisotropic lattices; see~\cite{aniso},
and for first one-loop results~\cite{anisoRen}),
there will be corrections to the weight factors of some of the loops at the 
boundary.
It is interesting to note that to obtain the counter term for 
$\Tr\, F_{0k}^2$ in the Wilson case by necessity affects the temporal 
weight factor of  plaquettes involving dynamical links (namely the  
temporal plaquettes touching the  boundaries at $x_0\s= 0$ and $x_0\s= T$),
whereas for the improved case one can choose to renormalize only the weight
factors of temporal plaquettes {\it within} the boundary
doubler layers, which do not involve dynamical links.
To obtain the 
counter term for $\Tr\, F_{kl}^2$ one can, in both cases, suitably
adjust the weight factor of spatial plaquettes within the 
boundary.\footnote{For the currently known one-loop corrections to the
boundary weights cf.~\cite{May} and references therein.}

This means that for the improved case some observables will not be sensitive
to $\Ord(a)$ quantum errors incurred by the wrong choice of the
temporal and spatial plaquette coefficients in the boundary; only observables
involving derivatives with respect to the boundary values that lead to
a coupling of the dynamical links with the boundary plaquettes will ``see''
such errors.

In any case, though,
for most applications of the \SF{}     it is      not important to
know the boundary counter terms.  The basic reason has already been
mentioned in the introduction: The boundary fields are used to probe the 
{\it local} dynamics of the theory, which, to leading order,     
should be independent of the precise details of the boundary terms, 
including their renormalization (see~\cite{May} for a formal proof).

\subsection{The Classical Background Field}

The boundary values of the gauge field
 induce some classical background field in the bulk.  It is important, 
e.g.~for perturbative calculations, that the background field be unique 
(up to gauge equivalence). This will not be the case for an arbitrary
choice of boundary values. 
Spatially constant boundary values have turned out to be both convenient and 
sufficient for all applications of the Schr\"odinger functional (so far, at 
least). Physically,
these boundary conditions correspond to a constant color-electric
background field. We will first consider generic boundary conditions of
this type and   later   discuss further restrictions that 
have to be imposed when we turn to the question of uniqueness.

In the {\it continuum} the background gauge field in such a situation,
denoted by $\bA_\mu(x)$, is 
given by
\beq
\bA_0(x) \= 0 ~, 
     \qquad \bA_k(x) \= {1\o T}\left [ x_0 C'_k + (T-x_0) C_k\right] ~.
\eeq
Here $C_k$ and $C'_k$ are the (hermitean) boundary values of the gauge field,
\beq
 \bA_k(x) ~=~ \left\{ 
\begin{array}{ll}
     C_k         & \quad \mbox{at $x_0=0$} \nn[1mm]
     C'_k        & \quad \mbox{at $x_0=T$} ~,
\end{array}  \right.
\eeq
in terms of which the color-electric field strength reads
$\bar{F}_{0k} = (C'_k - C_k)/T$.

If one assume that $C_k$ and $C'_k$ commute (e.g.~if they are diagonal, see
below), then an obvious guess for the background field on the {\it lattice}
is $\bU_\mu(x) = \exp(-i a_\mu \bA_\mu(x))$, or explicitly,
\beq\label{U_bkd}
  \bU_0(x) \= 1 ~, 
 \qquad  \bU_k(x) \= \exp\Bigl(-i {a_k\o T} [ x_0 C'_k + (T-x_0) C_k]\Bigr) ~.
\eeq
For the Wilson case this corresponds to the boundary values
\beq
  \bU_k(x) ~=~ \left\{ 
\begin{array}{ll}
   W_k(\x)  \= e^{-i a_k C_k}         & \qquad \mbox{at $x_0=0$} \nn[1mm]
   W'_k(\x) \= e^{-i a_k C'_k}        & \qquad \mbox{at $x_0=T$} ~.
\end{array}  \right.
\eeq
For improved actions one can read off the boundary fields $W_k^{(\pm)}$ and
$W'^{(\pm)}_k$ from~\eqn{U_bkd} at $t=-a_0,0$ and 
$t=T',T'+a_0$, respectively.
$W^{(0)}(\x) = 1 = W'^{(0)}(\x)$, of course.\footnote{Note
that we are reverse-engineering the boundary conditions: We start with
a gauge field that, as we will see, satisfies the lattice field equations,
and read off the boundary conditions from the boundary values of the 
field. This is perfectly legitimate.}

It remains to be proved that~\eqn{U_bkd} actually {\it is}
the correct background field on the lattice.
To this end, note that for this field the
non-vanishing contributions to the action come from the temporal
plaquette and rectangle terms,
\bea
 P_{0k}(x)\=P_{k0}(x)^\ast &=& \exp\Bigl(-i {a_0 a_k\o T}[C'_k -C_k]\Bigr)~,\nn
 R_{0k}(x)\=R_{k0}(x)^\ast &=& \exp\Bigl(-2i {a_0 a_k\o T}[C'_k -C_k]\Bigr)~.
\eea
The contribution of a loop to an arbitrary gauge action is simply
related to the color-electric flux through the loop. Since the background 
electric field is constant in space and time, the same is true for these fluxes,
which implies that~\eqn{U_bkd} obeys the lattice field equations for
{\it any} gauge action. 
The uniqueness of the background field~\eqn{U_bkd}
      will be discussed in the next subsection. 

The fact that~\eqn{U_bkd} satisfies the field equations for an arbitrary gauge
action,  testifies to the ``quality'' of the abelian
boundary fields, as previously remarked in~\cite{LNWW}.
This can also be seen by calculating the value of the plaquette $+$ rectangle
action~\eqn{PRactions} for the background field with \SF{} boundary conditions, 
\bea\label{bkd_action}
 S_g[\bU] &\s=& {T L_1 L_2 L_3 \o g^2} \, \sum_k \, \Tr\, 
 \Bigl( {2\o a_0 a_k} \sin\alpha_k \Bigr)^2 \,
 \Bigl( 1+ 4(c^{(1)}_k + c^{(1)}_0 ) \sin^2\alpha_k \Bigr) \nn
 &\s=& {L_1 L_2 L_3\o g^2 T} \, \sum_k \Tr \, (C'_k -C_k)^2  \+ \Ord(a^4) \: ,
\quad \alpha_k \equiv {a_0 a_k\o 2 T} (C'_k -C_k) ~.
\eea
The same result would have been obtained 
with periodic boundary conditions in the time direction.
We see that for constant abelian boundary fields even the Wilson
action has only $\Ord(a^4)$ errors relative to the continuum limit.

\subsection{Uniqueness of the Classical Background Field}

It is important to know if, for given boundary values of the gauge field,
the background field 
is the unique (up to gauge equivalence) absolute minimum of the action. 

In ref.~\cite{LNWW} a theorem establishing uniqueness was provided for the 
Wilson plaquette action, in any number of dimensions, for a certain class 
of diagonal and spatially constant boundary values.
To specify this class consider diagonal boundary values
\beq\label{C_k}
  C_k \= {1\o L_k} \pmatrix{ \phi_{k1} &   0       & \ldots & 0 \cr
                                 0     & \phi_{k2} & \ldots & 0 \cr
                             \vdots    & \vdots    & \ddots & \vdots\cr
                                 0     &   0       & \ldots & \phi_{kN} \cr }
\eeq
and similarly for $C'_k$ in terms of a set of $\phi'_{k\alpha}$.
Gauge symmetry implies that the $\phi_{k\alpha}$ and $\phi'_{k\alpha}$
have angular character; we can always choose them to lie in $(-\pi,\pi]$.
A vector $(\phi_1,\ldots,\phi_N)$ is said to lie
in the {\it fundamental domain} if
\beq\label{fund_dom}
 \sum_\alpha\, \phi_\alpha = 0 \, , \quad \phi_1 < \phi_2 < \ldots< \phi_N \, ,
     \quad \phi_N - \phi_1 < 2\pi \, .
\eeq
It was proved in~\cite{LNWW} that a sufficient condition for the
uniqueness of the background field~\eqn{U_bkd} is that the vectors
$\phi_k$, $\phi'_k$ (defined in the obvious way) lie in the fundamental
domain.\footnote{Another technical requirement in the proof is that the
space-time volume is not too small. The lower bound given in~\cite{LNWW}
reads $\,T L \geq 20\, a_0 a\,$ for SU(3) gauge fields on a spatially 
symmetric lattice.}
The theorem also holds if some or all of the $\phi_k$ or $\phi'_k$ are zero
(vectors).

The proof given in~\cite{LNWW} for the Wilson action is quite involved.
As a corollary it implies that the analogous theorem holds in the
continuum. One therefore expects such  a theorem to hold
for improved actions, since they are supposedly more continuum-like
than the Wilson action.  However, there is at least one step in the
proof, related to the use of the field equations, that prevents
a straightforward extension to improved actions.

Even though uniqueness presumably holds for all ``reasonable'' improved
actions, we have no proof at the moment. We have therefore investigated 
the question of uniqueness numerically. For various actions, volumes and
boundary conditions we started with some random 
initial configurations, and used simulated annealing to find a minimum
of the action. Besides the plaquette $+$ rectangle action, we considered
the action with an additional ``parallelogram'' term, i.e.~a set if loops
sufficient for full order $a^2$ improvement~\cite{LWGlue}. We stepped through
the region of improvement coefficients of interest in practice --- and beyond.
Several boundary values in (and on the boundary of) the fundamental domain
were considered.
In all cases where the coefficients of the 
improvement terms are such that positivity holds (cf.~\cite{LWGlue}; any
viable action must satisfy positivity),
the simulated annealing algorithm eventually converged to the classical
background field~\eqn{U_bkd}.  

We should add that we found the above to be true even on the very small
lattices, like $L\s= 2a$, $T\s= 3a$,\footnote{The case $T\s= 2a$ is trivial
for improved actions, since there are no dynamical spatial links then.
But we checked $T\s= 2a$ for the Wilson case.}
 where the proof of~\cite{LNWW}
does not apply. The only cases we could find where the global minimum is
definitely not given by~\eqn{U_bkd} involve actions with large negative
coefficients for the improvement terms, violating positivity. Then indeed
rather interesting things can happen. The simplest, and surprisingly rich,
example is provided by the  plaquette $+$ rectangle  action on a 
two-dimensional lattice with $L\s=2a$, $T\s= 3a$  and vanishing boundary
fields, $\phi = 0 = \phi'$. For a certain region of large negative
rectangle coefficient, for instance, there is a discrete set of 
degenerate vacua that break spatial translation symmetry. The vacua have
an anti-ferromagnetic character, in that translation invariance holds only 
under shifts by two lattice units.

We conclude that for boundary values in the fundamental domain all improved
actions of interest in practice seem to have the expected unique global
minimum for arbitrary volumes.

\clearpage

\section{Fermions}\label{sec:quarks}

\subsection{Fermion Actions in the Bulk}

Before writing down the most general quark action 
\beq\label{qrk_action}
 S_q[\psib,\psi,U] ~=~ 
    a_0 a^3 \, \sum_x \, \psib(x) \, Q \, \psi(x)
\eeq
we will be interested in, let us consider two important special cases 
on isotropic lattices. These are the ``isotropic SW''~\cite{SW} and the
``isotropic D234''~\cite{ILQA} actions, whose quark matrices $Q$
can be written as 
\bea\label{Wactions}
  Q_{W_1} &\s=& m_0 \- a \,{\om \o 4}\, \sigF \+ \sum_\mu \ga_\mu W_\mu ~, \nn
  Q_{W_2} &\s=& m_0 \- a \,{\om \o 6}\, \sigF \+ 
       \sum_\mu \, \Bigl( \ga_\mu W_\mu + {a\o 6} W_\mu^2 \Bigr) ~,
\eea
in terms of the ``Wilson operator''
\beq
 W_\mu ~\equiv~ \del_\mu - {a_\mu \over 2} \ga_\mu \De_\mu \: .
\eeq
We will therefore also refer to these actions as the W$_1$ and W$_2$
actions, respectively, as anticipated  in eqs.~\eqn{Wactions}.

In the above $\sigF$ is the so-called clover term (cf.~appendix~A), and
we used the standard first and second order covariant
derivative operators, defined, on a general lattice, as\footnote{We ignore
tadpole factors for the link fields that one might want to use in the
context of an actual simulation.}
\bea
 \del_\mu \psi(x) &\equiv& 
 {1\over 2a_\mu}\, \biggl[ U_\mu(x) \psi(x+\mu) - U_{-\mu}(x) 
           \psi(x-\mu)\biggr] \\ 
 \De_\mu \psi(x)  &\equiv& 
 {1\over a_\mu^2} \, \biggl[ U_\mu(x) \psi(x+\mu) + U_{-\mu}(x) \psi(x-\mu)
                                       -2 \psi(x) \biggr] 
\eea
(cf.~appendix~A for unexplained notation).
The main point of the Wilson operator is that it can be expressed in terms
of two-dimensional (spinor-) projection operators   as
\beq
  \ga_\mu W_\mu ~=~ -\del_\mu^{+} P_\mu^{-} + \del_\mu^{-} P_\mu^{+} \: ,
 \qquad\quad  P_\mu^{\pm} \equiv \half (1\pm \ga_\mu) \: ,
\eeq
where we   introduced forward and backward covariant derivatives,
\beq
\del_\mu^{\pm} \psi(x) ~\equiv~ \pm{1\over a_\mu} \biggl(
 U_{\pm \mu}(x) \psi(x\pm \mu) - \psi(x) \biggr) ~.
\eeq
This property allows for an efficient coding of the W$_1$ and W$_2$ actions.

On the classical level, where the clover coefficient $\om=1$,
the W$_1$ action has $a^2$, the W$_2$ action
$a^3$ errors, and they are the ``cheapest'' isotropic
actions with these errors. 
They are
natural candidates for $\Ord(a)$ quantum improvement, which involves
tuning the clover coefficient $\om$ non-perturbatively.

As we will see, to be able to define simple Dirichlet boundary conditions
all we need is
 that the temporal parts of the quark actions have the above projection 
properties.  On anisotropic 
lattices that are spatially isotropic, 
the actions of interest can therefore be written as
\beq\label{qrk_matrix}
 Q ~=~ \ga_0 \, W_0 \+ b_0 \, a_0 \, W_0^2 \+ c \, {\bf D}\sl \+ M \, .
\eeq
We have separated the terms in the quark matrix according to their
$\gamma$-matrix structure.
For the spatial part of~\eqn{qrk_matrix} we need to make no special
assumptions, so ${\bf D}\sl \equiv \sum_k \ga_k D_k$ is written in terms
of generic derivative operators 
$D_k = \del_k + \Ord(a_k^2)$,\footnote{Usually 
we use $D_\mu$ to denote the covariant continuum derivative, but its
use in this section as a generic lattice derivative should
not cause any confusion.}
and $M$ denotes all other terms,
\beq
 M ~=~ m_0 \- {r a_0 \o 2} \, 
   \biggl(\om_0 \sum_k \si_{0k} F_{0k} \+ \om \sum_{k<l} \si_{kl} F_{kl}
           \+ \sum_k \De_k + \ldots \biggr ) ~.
\eeq
The natural generalization of the W$_1$ and W$_2$ actions to anisotropic
lattices, which we will also refer to by these names, involves
\bea\label{br_values}
 \mbox{W$_1$ action:}  & b_0 = 0 \, ,       & r = 1 \nn[1mm]
 \mbox{W$_2$ action:}  & \displaystyle{b_0 = {1\o 6}} \, , & r = {2\o 3} \: ,
\eea
as on isotropic lattices.

As shown in appendix~C, quantum $\Ord(a)$ on-shell improvement of such
actions requires the tuning of three coefficients: The temporal and spatial 
clover coefficients, $\om_0$ and $\om$,  as well as $c$,
 the ``bare velocity of light'',
that has to be adjusted to obtain a renormalized velocity of light
of $1$.\footnote{In the case of dynamical simulations on anisotropic
lattices one might have to iteratively tune parameters in the glue and the 
quark actions until they are self-consistent, i.e.~correspond to the same
renormalized anisotropy.}

We should add a remark on notation. For the isotropic W$_1$, or SW, 
action $c_{{\rm SW}} \equiv r \, \om$ is often referred to as the 
clover coefficient. It should not cause confusion when, more generally,
we use that name for $\om$.

\subsection{Field Equations and Dirichlet Boundary Conditions}\label{sec:qrkDir}

Dirichlet boundary conditions for Wilson quarks were derived in~\cite{Sint}
using the transfer matrix formalism~\cite{LTM} (cf.~also~\cite{Smit}).
This is somewhat involved, especially for improved quark actions.
Since ultimately we do not need the transfer matrix itself,
we will use a shortcut.
Namely, a little bit of thought shows that to decide which field components 
can be consistently fixed at the boundary it is sufficient to consider the 
projector  structure of the quark action (or field  equations).

\begin{figure}[tp]
\vskip -40mm
\mbox{ \hspace{ 2 em}\ewxy{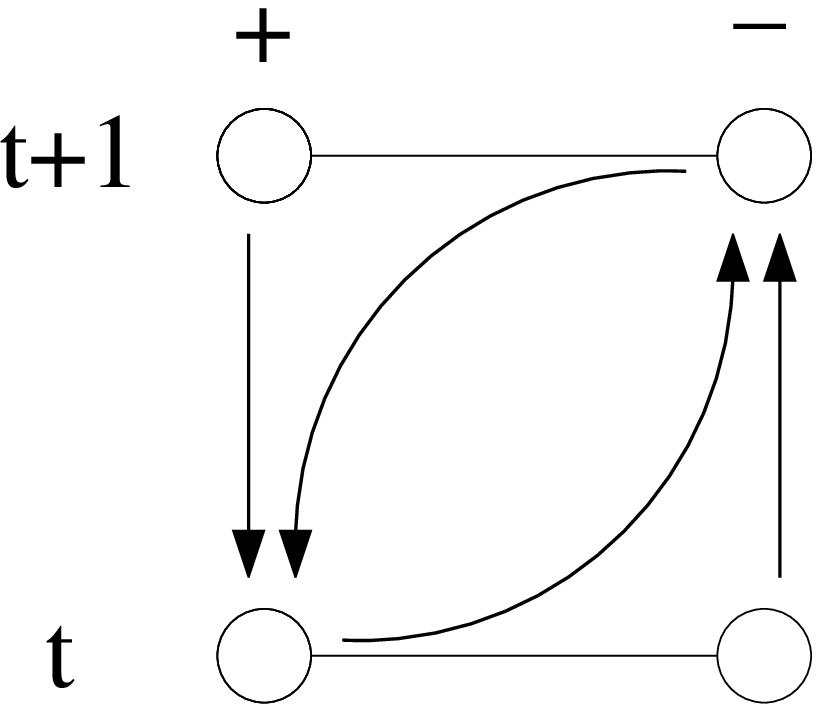}{110mm} 
       \hspace{-12em}\ewxy{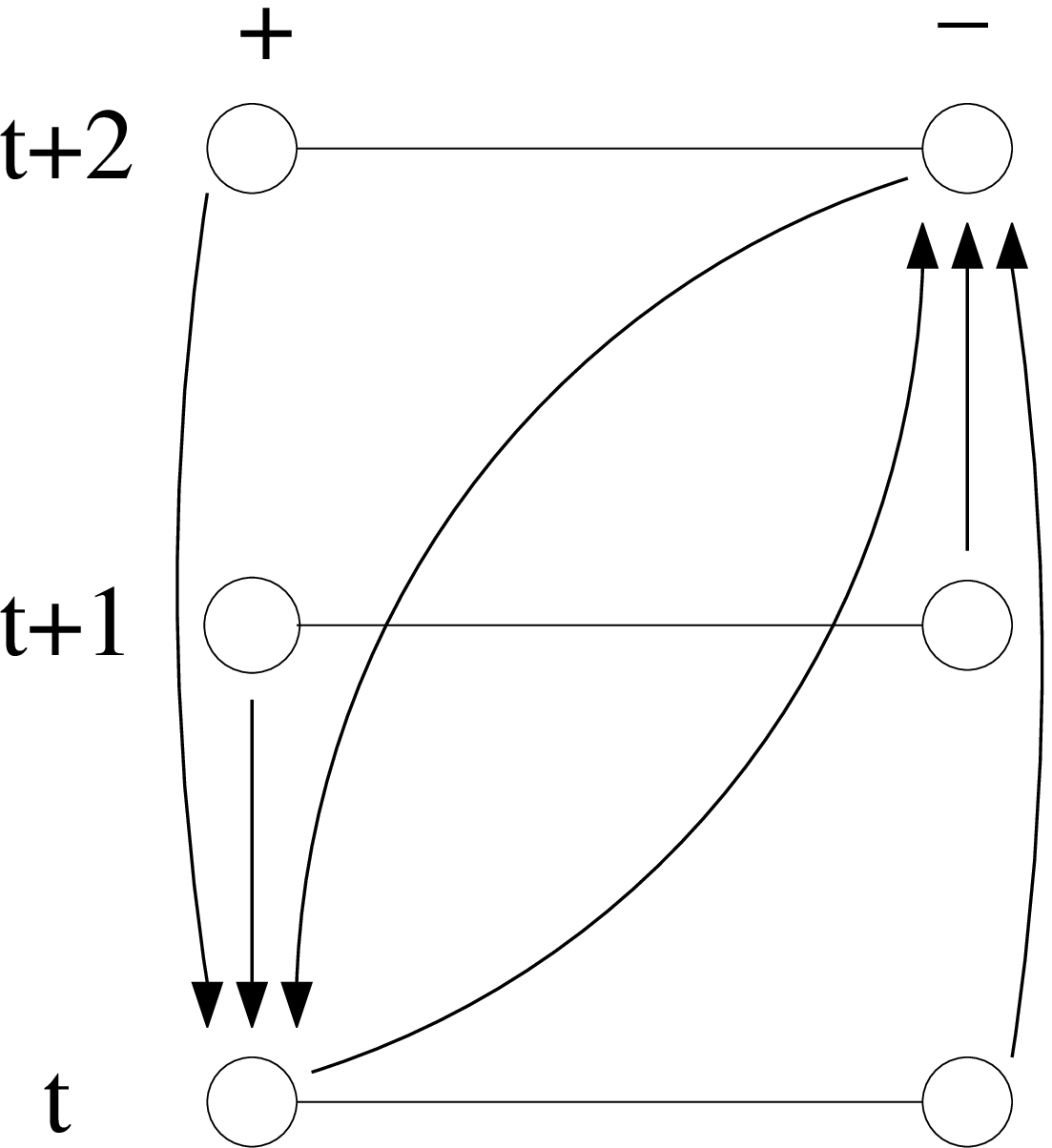}{110mm} } 
\vskip 16mm
\caption{
Dependence diagrams of $P^+\psi$ (left side of each figure) and $P^-\psi$
(right side of each figure) for W$_1$ (left) and
W$_2$ (right) equations of motion. 
The field components pointed to by the arrows are determined by the
field components in the time slice(s) at which the arrows originate. 
The corresponding diagrams for the $\psib P^\pm$ components are
obtained by reflection with respect to a vertical line through the
middle of each diagram (or by exchanging $+$ and $-$ at the top).
}
\label{fig:eom}
\vskip 8mm
\end{figure}

\begin{figure}[tbhp]
\vskip -18mm
\mbox{ \hspace{ 4 em}\ewxy{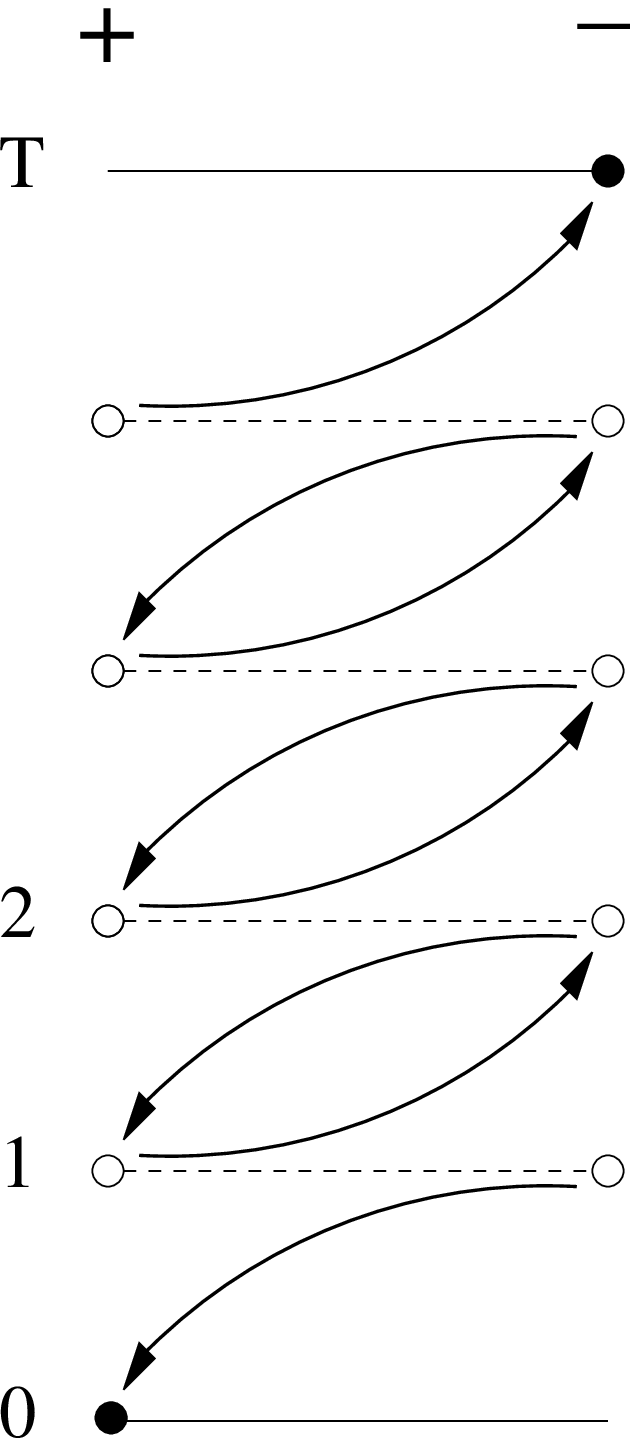}{130mm} 
       \hspace{-14em}\ewxy{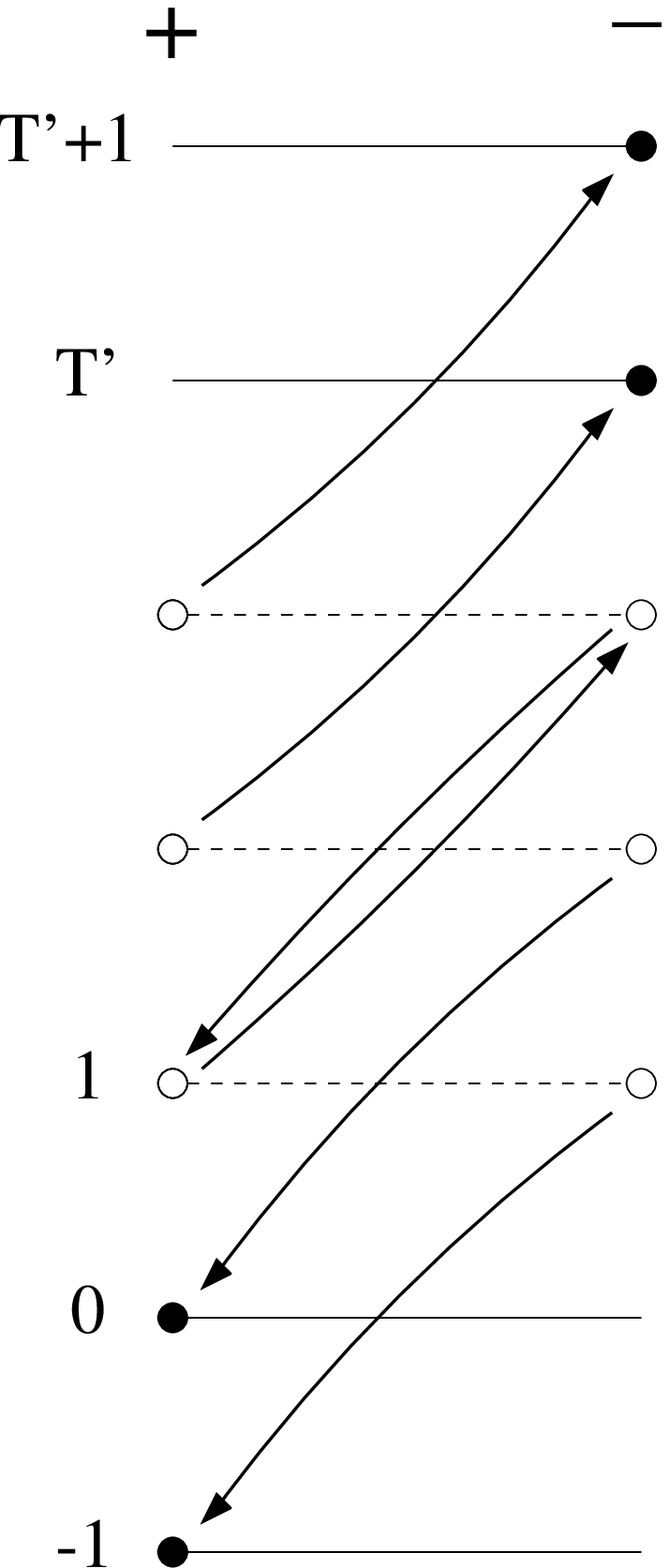}{130mm} } 
\vskip 16mm
\caption{Schematic representation of the fermion field equations with 
Dirichlet boundary conditions for a W$_1$ (left) and W$_2$ (right) action.
 For clarity the field equations are 
represented only by the ``diagonal'' arrow of figure~\protect\ref{fig:eom}; 
what matters is that one arrow corresponds to one equation for a field 
component in terms of field components on other time slices
(represented more correctly by two, respectively, 
three arrows in figure~\protect\ref{fig:eom}).
This implies that the a priori unknown field values in the bulk (open circles) 
are uniquely determined by the boundary values (solid circles) and the 
field equations; there are exactly as many unknowns as equations.
The corresponding diagrams for the $\psib P^\pm$ components are
obtained as in figure~\protect\ref{fig:eom}.}
\label{fig:eom_all}
\vskip 10mm
\end{figure}

The lattice equation of motion of $\psi$ in the bulk is derived by varying
the action defined by~\eqn{qrk_matrix} with respect to $\psib$. Projected 
onto $P^\pm \equiv P^\pm_0$ components the field equations read
\beq\label{eom}
 \Bigl[ \pm \del_0^\mp \+ b_0 \,a_0 \,(\del_0^\mp)^2 \Bigr ] P^\pm \psi(x) \= 
  - M^\pm P^\pm \psi(x) \- c \sum_k \ga_k D_k P^\mp \psi(x) ~,
\qquad  0 < x_0 < T' \, ,
\eeq
with $M^\pm \equiv P^\pm M P^\pm $.
Analogous equations are obtained for the $\psib$ components.
Diagrammatically the structure of these equations is illustrated in
figure~\ref{fig:eom}.
 When one now asks 
how to fix some field components at the boundary in a way that is
{\it consistent} --- leading to a system of equations that is neither 
under- nor overdetermined --- one finds the situation depicted 
in figure~\ref{fig:eom_all}.\footnote{In figures~\ref{fig:eom} 
and~\ref{fig:eom_all} we assume a {\it fixed} gauge field background. 
This is compatible with the standard method of performing the path integral 
in lattice gauge theory: first integrate
out the fermions analytically for fixed gauge field, then perform the path
integral over the latter by Monte Carlo. We mention this because otherwise 
the (standard form of the) clover term hidden in $M^\pm$ would lead to 
complications; it connects to time slices beyond the ones shown in 
figure~\ref{fig:eom}. Also, this is another reason to avoid using the
transfer matrix for a combined fermion and gauge system. Due to the clover
term it would in general be more complicated than simply the (symmetrized) 
product of the gauge and fermion transfer matrices. This lead the authors 
of~\cite{FNAL} to propose a modified clover term that extends only over two 
time slices instead of three.
}

This implies that for a W$_1$ action consistent Dirichlet boundary
conditions correspond to specifying
\bea\label{W_bc}
 P^+ \psi(x)       ~=&\rho(\x)   & \quad\mbox{at ~$x_0 = 0$} \nn
     \psib(x) P^-  ~=&\rhob(\x)  & \quad\mbox{at ~$x_0 = 0$} \nn
 P^- \psi(x)       ~=&\rho'(\x)  & \quad\mbox{at ~$x_0 = T'$} \nn
     \psib(x) P^+  ~=&\rhob'(\x) & \quad\mbox{at ~$x_0 = T'$} ~,
\eea
as derived in~\cite{Sint} using transfer matrix methods.
Note that it makes sense to couple W$_1$ quarks to either Wilson or
improved glue. 

W$_2$ quarks, on the other hand, can be consistently coupled only to improved
glue with Dirichlet \BCs; this will always be understood in the
following. In this case we have to specify not only the field components
indicated in~\eqn{W_bc} on the {\it inner} boundary layers, 
but the same components also on the {\it outer} boundary layers
at $x_0=-a_0$ and $T'+a_0$ (cf.~figure~\ref{fig:wgts}). 
To avoid cumbersome notation we will
not introduce additional symbols for the boundary values on the outer
boundary layers. It will always be clear from the context if
$\rho,\rhob$ etc refer to the inner or outer boundary, or both.

All components at the boundaries not of the general form shown in 
eq.~\eqn{W_bc}  have no influence on the bulk
dynamics and could in principle be set to arbitrary values. However,
for notational convenience we will set them to $0$, and 
formally define $\psi(x)$ and $\psib(x)$ for all $-\infty < x_0 < \infty$
by ``padding with zeros''. That is, we set $\psi(x)$ and $\psib(x)$ to 
zero for $x_0 < 0 $ and $x_0 > T$ in case of a W$_1$ action, and 
for $x_0 < -a_0 $ and $x_0 > T'+a_0$ in case of a W$_2$ action.
This has the advantage that we can now define the lattice quark
actions with Dirichlet boundary conditions to be given by exactly the
same sum as for $T=\infty$, with the boundary
values as discussed above.

Actually, we have cheated here. All we have shown above is which field
components are to be fixed at the boundaries. The question is, however,
whether the boundary terms in the fermion action need to be
given  ``special'' coefficients to avoid boundary errors (as we have seen
for the gauge action). It turns out 
that this is not the case for the W$_1$ action~\cite{LWtree}.
We will briefly discuss this issue later. For the moment, 
we will proceed by assuming that
the W$_1$ and W$_2$ actions with Dirichlet \BCs{} defined in the
previous paragraph only have the $\Ord(a^n)$ on-shell errors of the
bulk actions (with $n\s= 2$ for W$_1$, $n\s= 3$ for W$_2$).

One of the important features of the \SF{} for practical applications
is that with Dirichlet \BCs{} there are no zero modes at vanishing
quark mass, at least classically. This is true even in the continuum
limit with {\it homogeneous} Dirichlet boundary conditions,
$\rho \s= \rhob \s= \rho' \s= \rhob' \s= 0$, where, at tree 
level, the lowest eigenvalue of the Dirac operator is
$\pi/2T$ (one actually has to consider eigenvalues of $\ga_5 D\sl$; 
see~\cite{Sint,SinSom}  for details).

\subsection{Boundary Fields}

It is very useful (e.g. in the non-perturbative determination
of the clover coefficient, see sect.~\ref{sec:PCAC}) 
to define {\it boundary fields}~\cite{May}
as functional derivatives with respect to the boundary values of the quark
fields. 

For the case of a W$_1$ action these functional derivatives
will always be with respect to the boundary values in eq.~\eqn{W_bc}.
To see explicitly what these boundary fields look like, consider the
terms in a W$_1$ action coupling to the (inner) boundary at $x_0=0$, say,
\beq\label{Sbdy}
 a^3 \sum_{\x} \, 
 \Biggl[ \rhob(\x) \, a_0 \Dssl \, \rho(\x) 
         \- \rhob(\x) U_0(0,\x) P^- \psi(a_0,\x) 
           \- \psib(a_0,\x) P^+ U_0(0,\x)^\dag \rho(\x) \Biggr] \, .
\eeq
There is a similar formula for the terms coupling to the boundary at
$x_0 \s= T$.
The correlation functions of the boundary fields
we will consider later only involve the zero momentum Fourier 
components of the boundary fields. The spatial derivative term in~\eqn{Sbdy}
does not contribute then, and we will from now on drop such terms.  

Let us define the boundary fields by applying
\bea\label{zeta_W_def}
 \zeta(\x)       ~\equiv&
            \displaystyle{{\delta\o \delta \rhob(\x)}} ~, \quad
 \bar\zeta(\x) ~\equiv\! &-{\delta\o \delta \rho(\x)}  \nn
 \zeta'(\x)      ~\equiv&
            \displaystyle{{\delta\o \delta \rhob'(\x)}} ~, \quad
 \bar\zeta'(\x)  ~\equiv& -{\delta\o \delta \rho'(\x)}  \:
\eea
within the functional integral (as left-derivatives, say)
and then setting all $\rho$ and $\rhob$ to zero.
Since not all components of the boundary values are independent
(e.g. $P^-\rho \equiv 0$ from eq.~\eqn{W_bc}), the above definition is 
ambiguous. To achieve uniqueness we impose the natural constraints
\bea\label{unique}
 P^+ \, \zeta(\x)      ~=&\! 0 \, , \quad
 \bar\zeta(\x) \, P^-  ~=&0            \nn
 P^- \zeta'(\x)     ~=&0 \, , \quad
 \bar\zeta'(\x) P^+ ~=&0 \: .
\eea
Explicitly, we then find
\bea\label{zeta_expl}
 \zeta(\x)      &=&   U_0(0,\x)  \,       P^- \, \psi(a_0,\x) \nn[0.6mm]
 \bar\zeta(\x)  &=& \psib(a_0,\x)  \,     P^+   \, U_0(0,\x)^\dag \nn[0.6mm]
 \zeta'(\x)     &=&  U_0(T'\!-\!a_0,\x)^\dag\, P^+ \,
                                          \psi(T'\!-\!a_0,\x)\nn[0.6mm]
 \bar\zeta'(\x) &=& \psib(T'\!-\!a_0,\x)   \,   P^- \, U_0(T'\!-\!a_0,\x) \: .
\eea

For a W$_2$ action one can in principle define two sets of boundary
fields, either by taking derivatives with respect to the inner, or 
with respect to the
outer boundary values. However, the latter are much simpler: 
The terms in the action coupling to the outer boundaries are given by
(ignoring spatial derivative terms)
\bea\label{Sbdy_Wtwo}
&& \hspace{-9mm} b_0 \, a^3 \sum_{\x} \, 
 \Bigl[
 \rhob(\x)\, U_0(0,\x) \,P^- \, \psi(a_0,\x) 
          \+ \psib(a_0,\x)\, P^+ \, U_0(0,\x)^\dag \, \rho(\x) \nn
&& \hspace{-9mm}
 \+ \rhob'(\x) \, U_0(T'\sm a_0,\x)^\dag \, P^+ \, \psi(T'\sm a_0,\x) 
 \+ \psib(T'\sm a_0,\x) \,P^- \,U_0(T'\sm a_0,\x) \,\rho'(\x) \Bigr] \,, \quad
\eea
where we have used the fact that the temporal links {\it within} the double
boundary layers are all $1$. 
We see that if we now define, with respect to the outer boundary values,
\bea\label{zeta_Wtwo_def}
 \zeta(\x)       ~\equiv&
        \displaystyle{ -{\delta\o b_0\delta \rhob(\x)} } ~, \quad
 \bar\zeta(\x) \,~\equiv& {\delta\o b_0\delta \rho(\x)}  \nn
 \zeta'(\x)    \,~ \equiv&     
         \displaystyle{-{\delta\o b_0\delta \rhob'(\x)} } ~, \quad
 \bar\zeta'(\x)  ~\equiv& {\delta\o b_0\delta \rho'(\x)}  
\eea
the explicit expressions for the boundary fields are 
identical to those for W$_1$ actions, eq.~\eqn{zeta_expl}!

\subsection{Quark Propagator with Dirichlet Boundary Conditions}
 
The correlations functions we are interested in correspond to 
homogeneous Dirichlet \BCs{},
 $\rho\s= \rhob\s= \rho' \s= \rhob' \s= 0$ (remember that
for the boundary fields we set the $\rho$'s to zero after taking the
functional derivatives).
The propagator 
\beq G(x,y) \= [ \psi(x) \psib(y) ]_F
\eeq
 in a given gauge background ($[\ldots ]_F$ denotes a fermionic contraction)
therefore satisfies
\beq\label{prop}
 \sum_z Q(x,z) \cdot G(z,y) \= {1\o a_0 a^3} \, \delta_{xy} ~,
  \qquad\quad 0 < x_0, y_0 < T' \, ,
\eeq
with the boundary conditions that $G(x,y)$ vanishes when $x$ or $y$ are
on either boundary.
Eq.~\eqn{prop} can be solved by standard iterative methods.

It is easy to check that with homogeneous Dirichlet boundary
conditions $\ga_5 Q$ is hermitean (as long as $M$ and $D_k$ in
eq.~\eqn{qrk_matrix} are hermitean, respectively, anti-hermitean),
so that
\beq\label{Gdag}
   G(x,y)^\dag \= \ga_5 \, G(y,x) \, \ga_5 \: ,
\eeq
as for the usual case of periodic boundary conditions in all directions.

\vskip 1mm
We end this section with some brief remarks on the boundary errors
of the fermionic part of the action. A detailed discussion will not
be presented here, but we would like to point out the difference to
the gluonic boundary errors.

The main point is that Wilson-type quark actions are only
{\it on-shell} improved, even classically and with periodic boundary
conditions. This is due to the field redefinition necessary
to avoid the doubler problem (cf.~appendix~C). 
One strategy for the determination of the coefficients of the 
boundary terms in the quark action is the study of correlations
functions.   Namely, to ask for which boundary coefficients 
the correlation functions of bulk fields, $\psi(x)$ and $\psib(x)$, as
well as boundary fields, $\zeta(\x)$ etc,  are improved after
appropriate field renormalizations have been performed.

At tree-level it is sufficient to consider all possible two-point
functions, by Wick's theorem.  It is not difficult to see that
with homogeneous boundary conditions the bulk field 
renormalization is given by the usual field redefinition factor,
cf.~appendix~C.
This is true for any value of the boundary coefficients in the
action (since the bulk fields vanish at the boundary).
The latter {\it are} constrained by the demand that the 
{\it boundary fields}  renormalize multiplicatively.

A detailed discussion of the two-point functions involving
boundary fields can be found in~\cite{LWtree} for the W$_1$ case.
The conclusion is that the coefficients of the boundary terms in the
action should be exactly as in the bulk (cf.~subsect.~\ref{sec:qrkDir}); 
then the boundary fields are improved once they are renormalized by the 
{\it inverse} factor of the fields in the bulk.

For the W$_2$ case the same is true for the zero momentum components of
the boundary fields. When considering general boundary fields one has
to appropriately adjust the coefficients of spatial derivative terms
at the boundary (which we ignored, except in eq.~\eqn{Sbdy}, since
they are irrelevant for the applications we have in mind).

On the quantum level the general discussion of boundary terms at 
$\Ord(a)$ is identical for W$_1$ and W$_2$ actions. We refer to~\cite{May}
for the details. Of immediate interest to us is only the conclusion
that the quantum boundary effects do not matter for the applications
we are interested in. 

\section{PCAC and Chiral Symmetry Restoration}\label{sec:PCAC}

Here we sketch one of the simplest but most important
application of the \SF, namely the determination of the non-perturbative 
clover coefficient(s).
Though much of the following is simply lifted  from~\cite{Sep}, we
think it is at least pedagogically useful to go through this exercise, 
demonstrating, for example, how to maintain classical (plus tadpole, perhaps)
improvement to higher order, while tuning quantum errors at $\Ord(a)$.
This includes some discussion of the anisotropic case.
We will also provide a comparison of the W$_1$ and W$_2$ actions at
tree-level.

Consider lattice QCD with (at least) two flavors of mass-degenerate quarks.
In the continuum limit we expect the isovector axial current
$A_\mu^a$ to satisfy  the PCAC relation 
\beq\label{PCAC}
     \p_\mu A_\mu^a \= 2m P^a \: ,
\eeq
with the associated pseudo-scalar density $P^a$, as long as the quark
mass $m$ is sufficiently small. The idea behind using the PCAC relation
in restoring chiral symmetry non-perturbatively, is the following.
The PCAC relation is a  local  relation between operators
(imagining, momentarily, that we continued to Minkowski space).
It should
therefore hold with the {\it same mass} $m$ independent of what global
boundary conditions are used,\footnote{Note in particular, that the 
renormalization of the boundary terms in the gauge and quark actions has
no effect~\cite{May}, as already remarked in sect.~2.
One can therefore use the classically improved \SF{}
when determining the clover coefficient non-perturbatively.} 
or what matrix elements of the PCAC relation are considered. 
As it turns out~\cite{Dec95}, if one uses the wrong
clover coefficient in the lattice action, $m$ has a significant
dependence on the kinematical parameters used to probe the theory. Requiring
the dependence on kinematical parameters to vanish provides a powerful
non-perturbative tuning method.

To make these ideas more precise one has to consider the 
improvement and renormalization of the the axial current and density
on the lattice (which can be anisotropic, at least for the moment).
Naive, unimproved expressions for $A_\mu^a$ and $P^a$ are given by 
\bea\label{AP}
   A^a_\mu(x) &=& \psib(x) \ga_\mu \ga_5 \half \tau^a \psi(x) \:, \nn
   P^a(x)     &=& \psib(x)         \ga_5 \half \tau^a \psi(x) \: .
\eea
Here $\tau^a$ is a Pauli matrix acting on the (suppressed) flavor
indices of the quark fields.\footnote{We normalize the Pauli matrices
such that $\Tr\, \tau^a \tau^b = 2 \delta_{ab}$.}

As shown in appendix~C, on the classical level improvement of the
operators $A^a_\mu(x)$ and $P^a(x)$ changes them only by the same overall
mass-dependent factor, to the order we are working. The PCAC {\it relation} is
therefore not affected.

On the quantum level there will be additional errors starting at $\Ord(a)$.
One can show that the only modification of the 
PCAC relation required at $\Ord(a)$ is, in continuum notation,
 the addition of a counter term  
\beq\label{counterA}
(\delta A^a_0 \, , \, \delta A^a_k ) \=
 (a_0 \, c_{A0} \, \p_0 P^a,  \, a \, c_A \, \p_k P^a)
\eeq
to the axial current $A^a_\mu(x)$\footnote{In
the W$_2$ case an improved discretization of the derivative 
in~\eqn{counterA}  has to be  used on the lattice, which, 
borrowing notation from covariant lattice derivatives, we can
write as $\del_\mu (1 - {1\o 6} a_\mu^2 \Delta_\mu)$.}
(see~\cite{May} for a thorough discussion of the renormalization of the 
PCAC relation).
Here we are ignoring the multiplicative renormalization of 
$A_\mu^a + \delta A_\mu^a$ and $P^a$ on the quantum level.
For the non-perturbative
determination of the clover coefficient the precise
values of these renormalization factors are not required. We will absorb 
them in the unrenormalized current
quark mass $m$ (and continue calling it $m$).

Consider now correlation functions of the PCAC relation with some product
of fields $\O$. Then
\beq\label{PCACcorrfct}
\del_\mu \, \< \, (A_\mu^a + \delta A_\mu^a)(x) \, \O \, \> \=
 2 m \, \< \, P^a(x) \, \O \, \> \+ \Ord(a^2) \, ,
\eeq
{\it if} we have chosen the coefficients of the improvement terms equal to 
their non-perturbative values (for the W$_2$ case the lattice derivative
in this and other equations to follow is understood to be the improved
one; cf.~the last footnote).
Note that $\O$ does not have to be a renormalized operator; the 
correction terms will cancel between the two sides of eq.~\eqn{PCACcorrfct}.
What is important, however, is that the support of $\O$ does not contain 
$x$, to avoid contact terms. Finally, we should remark that for the 
W$_2$ case the $a^2$ errors in~\eqn{PCACcorrfct} are pure quantum errors 
(for suitably defined $m$, cf.~above).

A good choice of $\O$ is given by a product of zero momentum quark
boundary fields~\cite{May,Sep}. Define, for
$X^a=A_0^a$ and $X^a=P^a$,
\beq\label{fone}
  f_X(x_0) ~\equiv~ -a^6 \sum_{a,\y,\z} \, \third\,
   \< \, X^a(x) \, \bar\zeta(\y)\, \ga_5 \half\tau^a \zeta(\z) \, \> \:,
\eeq
in terms of the lower boundary fields $\zeta$ and $\bar\zeta$.  
Due to spatial translation invariance the lhs depends only
on $x_0$, as anticipated by our notation (in practice one
should average over spatial $\x$ on the rhs to increase statistics).
We now see that the PCAC relation implies for the current quark mass,
\beq\label{moft}
 m(x_0) ~\equiv~ r(x_0) \+ a_0 \, c_{A0} \, s(x_0) \= m + \Ord(a^2) \, ,
\eeq
where
\beq\label{rs}
 r(x_0) ~\equiv~ {\del_0   f_A(x_0)\o 2 f_P(x_0)} \, , \qquad
 s(x_0) ~\equiv~ {\Delta_0 f_P(x_0)\o 2 f_P(x_0)} \, .
\eeq
Note that due to spatial translation invariance the dependence on $c_A$
has dropped out of eq.~\eqn{moft}.

So far we have considered the general case of anisotropic lattices.
The isotropic case is significantly simpler in practice, 
since then only $\om\equiv \om_0$ has to be tuned in the quark action.
In the anisotropic case the color-{\it electric} background field chosen
here should essentially couple only to the temporal clover
term, $\om_0 \, \si_{0k} \, F_{0k}$, and allow its non-perturbative
tuning, even if the spatial clover coefficient $\om$ is not
exactly at its correct value. In a second step one could tune
$\om$ by choosing boundary conditions corresponding to a 
color-{\it magnetic} background field (if necessary, one could then perform
a second iteration of tuning $\om_0$). 

With spatially periodic \BCs{}
a background magnetic field is necessarily quantized with a minimum
non-zero value that might well be too large (cf.~e.g.~\cite{Witt} and
references therein) for our purposes. To avoid this problem one might
want to choose suitable {\it fixed} \BCs{} in one (or more) of the
spatial directions, which allows one to generate an arbitrarily small 
(classical) background color-magnetic field. In other words, one
basically exchanges the time direction with one of the spatial
directions in the \SF.

Note that anisotropic lattices have a distinct advantage in the context
of \SF{} simulations: One can use a small temporal extent $T$ in physical 
units to give a large lowest eigenvalue (and thereby, presumably, alleviate 
problems with exceptional configurations), and still have a large number, 
$T/a_0$, of time slices required from a practical point of view.

These ideas should be subjected to some numerical studies (in the free
and interacting cases). Since we have
not done so yet, we will in the remainder of this section
restrict ourselves to isotropic lattices.

Let us continue describing the improvement conditions for determining the
clover coefficient $\om$. To eliminate  $c_A \equiv c_{A0}$  
from~\eqn{moft} one can consider the analogously defined $m'(x_0)$,
$r'(x_0)$ and $s'(x_0)$ in
terms of the upper boundary fields $\zeta'$ and $\bar\zeta'$.
Namely, set
\bea\label{fT}
  f'_A(T'\sm x_0) &\equiv& +a^6 \sum_{a,\y,\z} \, \third\,
   \< \, A_0^a(x) \, \bar\zeta'(\y)\, \ga_5 \half\tau^a \zeta'(\z) \, \> \: \nn
  f'_P(T'\sm x_0) &\equiv& -a^6 \sum_{a,\y,\z} \, \third\,
   \< \, P^a(x) \, \bar\zeta'(\y)\, \ga_5 \half\tau^a \zeta'(\z) \, \> \:
\eea
and define $r'(x_0)$ and $s'(x_0)$ exactly as in~\eqn{rs} with
$f_A(x_0)$, $f_P(x_0)$ replaced by $f'_A(x_0)$, $f'_P(x_0)$, respectively.
The equations for $m(x_0)$ and $m'(x_0)$ now imply that
\beq\label{cA}
\hat{c}_A(x_0,y_0) ~\equiv~
         -{1\o a} \,\, {r(x_0)- r'(y_0)\o s(x_0) - s'(y_0)}
\eeq
is an estimator (in the sense of probability theory) of $c_A$, up to
$\Ord(a)$ errors. 

Before proceeding we have
to specify the \BCs. For the W$_1$ case a good choice of ``boundary
angles'' (cf.~\eqn{C_k}) for the
determination of the clover coefficient was found to be~\cite{Sep}
\bea\label{phi_csw}
(\phi_1, \phi_2, \phi_3)  &=& 
 \bigl(-\sixth \, ,\,  0\, ,\,      \sixth \bigr) \, \pi \nn
(\phi'_1,\phi'_2,\phi'_3) &=& 
 \bigl(-\fivesixth \, ,\, \third \, ,\, \half \bigr) \, \pi
\eea
with vanishing quark ``twist'', $\theta_k=0$ (cf.~appendix~A).
With these values the background field strength $F_{0k}$ 
is still small in lattice
units, for typical $L$ and $T$ used in practice, but large enough to
provide a significant lever arm for the tuning of $\om$.
It is crucial that the boundary conditions are different at the upper
and lower boundary; otherwise one could not use eq.~\eqn{cA} to
eliminate $c_A$ from the problem of the $\om$ determination.

With the above boundary values it turns out that to keep the denominator 
in~\eqn{cA} as far away from zero as possible, it is advisable
to use small $x_0$ and $y_0$. They should not be too small, of course,
to avoid boundary effects. After some study, the choice $x_0=y_0=\fourth T$
was suggested~\cite{Sep} for the W$_1$ case on an 
$T\cdot L^3 = 16 \cdot 8^3$ lattice.

Let us now introduce the following observables as estimators of the 
(unrenormalized) quark mass,
\bea
 M(x_0, y_0)  &\equiv& r(x_0)  \+ a \, \hat{c}_A(y_0,y_0) \, s(x_0) \nn
 M'(x_0, y_0) &\equiv& r'(x_0) \+ a \, \hat{c}_A(y_0,y_0) \, s'(x_0) \, .
\eea
One can then employ the demand that
$\Delta M(x_0,y_0) \equiv  M(x_0, y_0) - M'(x_0, y_0)$ vanish,
for suitably chosen $x_0$ and $y_0$, as an improvement condition
to determine $\om$~\cite{Sep}.
Since $\Delta M(x_0,x_0)\equiv 0$, one should try to separate $x_0$ and $y_0$
as much as possible. Using $\Delta M(\threefourth T,\fourth T)$ 
is a good choice for the W$_1$ case.

\vskip 1mm
We should mention a few more details:
\begin{itemize}

\item  The tuning of $\om$ should ideally
be performed at zero quark mass. Within the $a^2$ ambiguity of the quark mass
one could use any $M(x_0,y_0)$ or $M'(x_0,y_0)$ to define the quark mass.
A good choice for the W$_1$ case is  $M\equiv M(\half T,\fourth T)$.

\item In practice it turns out that $\Delta M$ has only a very weak 
dependence on $M$. This is true in the free case for both the W$_1$
and W$_2$ actions up to quark masses as large as $aM\s=0.1$ or $0.2$.
In~\cite{Sep} this was also found to hold for the
interacting case (although only much smaller quark masses were investigated). 
One can therefore tune $\om$ at small
positive values of the quark mass, which alleviates problems with 
accidental quark zero modes on coarse lattices.
It remains to be investigated exactly how large a mass one can use in the
interacting case (cf.~the end of this section for some relevant results).

\item $\Delta M$ does not vanish in finite volume,
even at tree-level and zero pole mass. As improvement condition one should
therefore not demand that $\Delta M$ vanish, but rather that it equals 
its tree-level value $\Delta M^{(0)}$. This is a relatively small correction,
but should be taken into account, nevertheless,  because it 
guarantees that tuning at positive quark 
masses does  not lead to additional tree-level errors in $\om$
(note that in such a case $\Delta M^{(0)}$ has to be evaluated at a mass
that is correct at least at tree-level).

\item  Due to the higher derivatives present in a W$_2$ quark action one
expects the boundary effects to extend one more lattice spacing into the
bulk. For the W$_2$ case one should therefore presumably move $x_0$ and
$y_0$ in the improvement condition 
$\Delta M(x_0,y_0) = \Delta M^{(0)}(x_0,y_0)$  slightly further away 
from the boundary than in the W$_1$ case. The necessity to do so can 
already be seen in the free case; the exact values to use for 
$x_0$ and $y_0$ should be
fixed after some numerical studies have been performed.

\end{itemize}

To determine $\om$ 
one uses           the estimator~\eqn{cA} to eliminate $c_A$
from the equations for $m(x_0)$ and $m'(x_0)$. This is adequate for
the $\om$ calculation, but for a more precise determination of $c_A$
itself it is better~\cite{Sep} to compare correlation functions with
quark twist $\theta_k \neq 0$ with ones where $\theta_k \s=0$, in
both cases with vanishing background field, $\phi_k \s= \phi'_k \s= 0$. 
More precisely, one should equate the corresponding mass difference 
to its tree-level value. It was found~\cite{Sep} that this mass difference
is again only  weakly dependent on the quark mass (at least
once the improved action with the previously determined clover
coefficient is used).

\begin{figure}[bp]
\vskip 12mm
\centerline{
\mbox{ \hspace{ 1 em}\ewxy{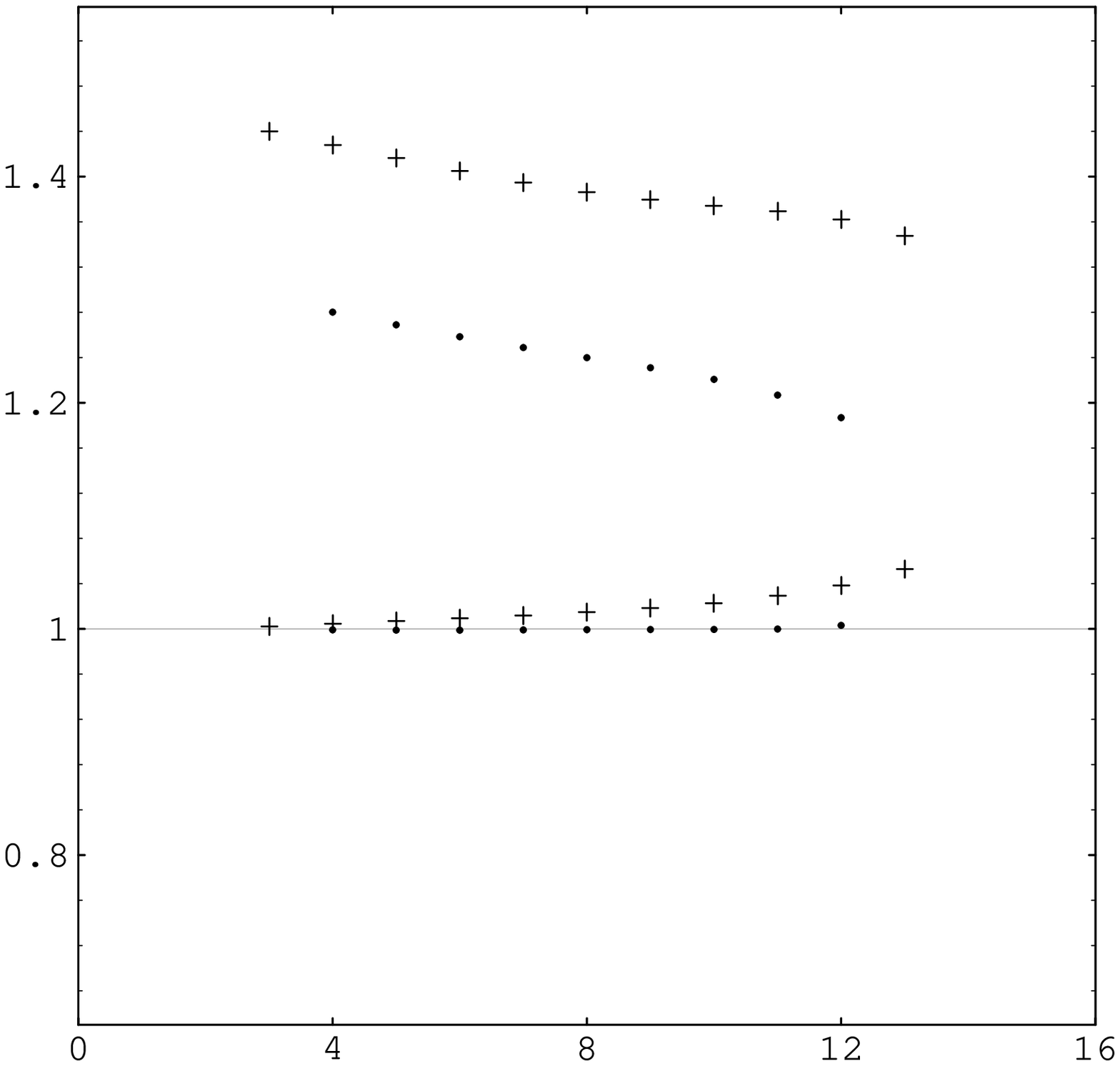}{90mm} 
       \hspace{-3 em}\ewxy{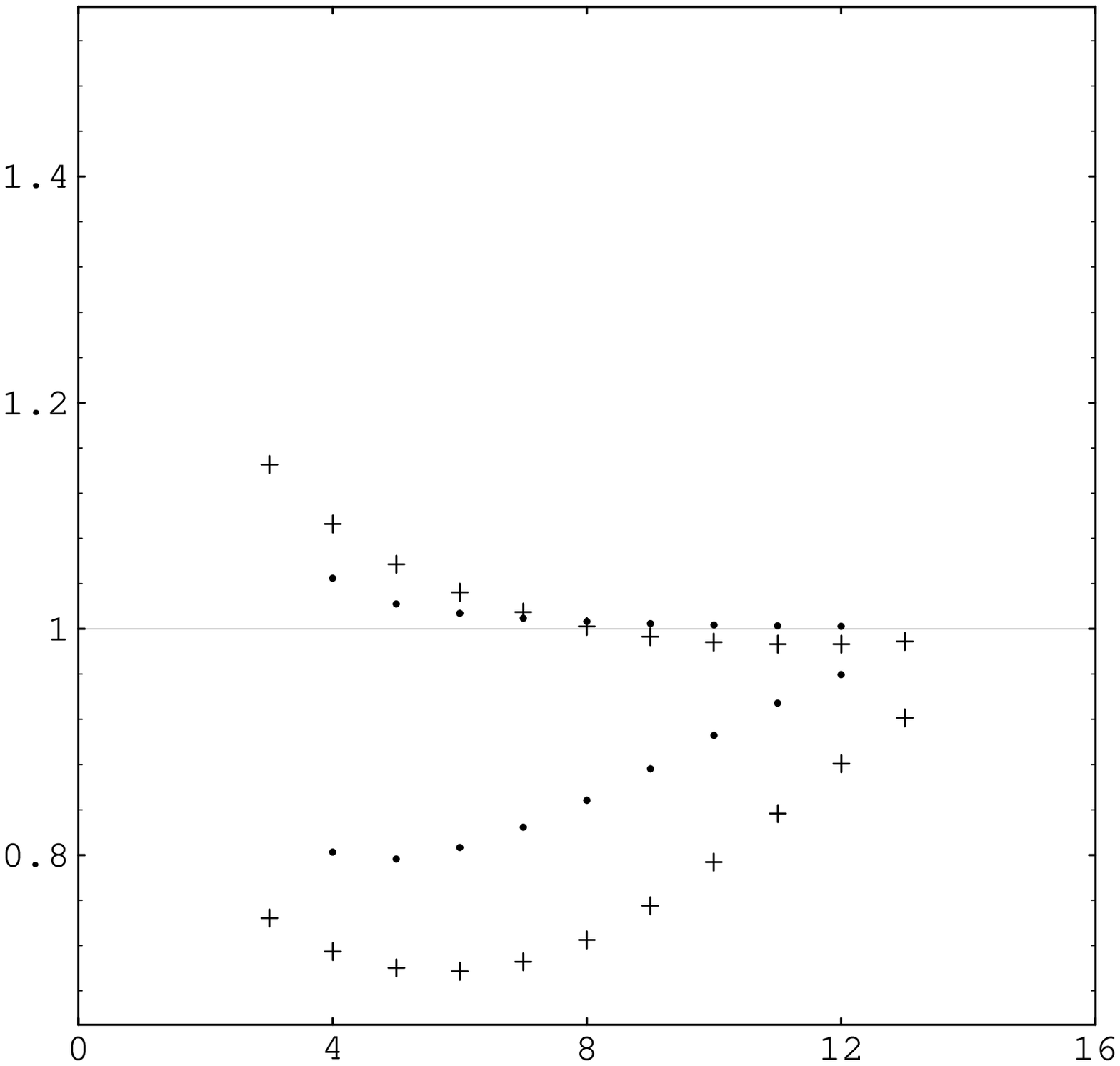}{90mm} } 
}
\vskip -9mm
\caption{The dimensionless ratios $r(x_0)/m_p$ (left) and 
$r'(x_0)/m_p$ (right) as a function of $x_0/a$,
for the free SW (or $W_1$, $+$) and D234 (or $W_2$, $\bullet$) actions on
an isotropic lattice.
$m_p$ is the pole mass, the zero momentum energy of a free lattice quark in
in the absence of a background field.
The points in the upper/lower half of the left/right figure 
correspond to vanishing clover coefficient $\om\s=0$;
the others to classical improvement, $\om\s=  1$.
The kinematical parameters used are, for both actions,
$a m_c \s= 0.02$, $T\s= 16a$, $L\s= 8a$,  $\theta_k \s= 0$,
with the angles $\phi, \phi'$ specifying the background field given by
eq.~\protect\eqn{phi_csw}.
}
\label{fig:r}
\vskip 2mm
\end{figure}

\begin{figure}[hbtp]
\vskip 2mm
\centerline{\ewxy{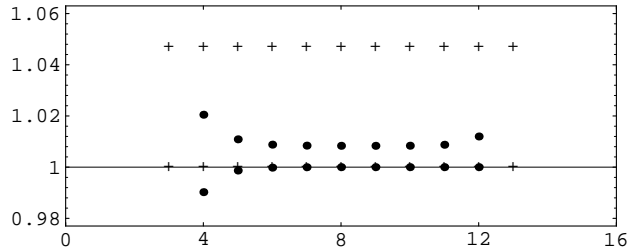}{110mm}}    
\vskip -38mm
\caption{As in figure~\protect\ref{fig:r}, but with 
vanishing background field, $\phi\s= \phi' \s= 0$. 
Now the points in the upper half correspond to $\theta_k \s=\pi/3$, 
the rest to $\theta_k \s= 0$ (in the absence of a background field
$r(x_0)=r'(x_0)$, and the value of $\om$ does not matter at tree-level).
}
\label{fig:th}
\vskip 2mm
\end{figure}

To give the reader some feeling for the various quantities involved in the
improvement conditions for $\om$ and $c_A$ we plot the 
tree-level     current quark masses with the relevant boundary
conditions in figures~\ref{fig:r} and~\ref{fig:th}, for both the
W$_1$ and W$_2$ actions. Recall that at tree-level
$c_A\s=0$, so that  $m(x_0)\s= r(x_0)$  and $m'(x_0)\s= r'(x_0)$.

We should comment on one aspect of figure~\ref{fig:r}.
Note that for the  W$_2$ action with $\om=0$ the current
quark mass estimates $r(x_0)$ and $r'(x_0)$ deviate less from the
pole mass than for the corresponding W$_1$ case.
The reason for this is simply the smaller value of the 
Wilson parameter ($r\s= 2/3$ versus $r\s= 1$), making the W$_2$ action
somewhat less sensitive to errors involving the background field.
This suggests to use a slightly larger background field when tuning
a W$_2$ action.

For completeness we end this section by describing  how to calculate the
correlation functions $f_A(x_0)$ and $f_P(x_0)$ in practice (analogous
remarks hold for $f'_A(x_0)$ and $f'_P(x_0)$).

Writing the operator $X$ in the form
 $\psib(x)\, \half \tau^a \Gamma_X \, \psi(x)$, we have 
\bea\label{ftwo}
 f_X(x_0) &=& -a^6 \sum_{a,\y,\z} \, \third\,
     \< \, \psib(x) \, \half \tau^a\Gamma_X \, \psi(x) \, 
               \bar\zeta(\y)\, \ga_5 \half\tau^a\zeta(\z) \, \> \nn
 &=& +\half a^6 \sum_{\y,\z} \, \, 
 \< \, \Tr\, \ga_5 \, [\zeta(\z)\psib(x)]_F \,\Gamma_X \,
                        [\psi(x)\bar\zeta(\y)]_F \, \>_G \: ,
\eea
where $\<\ldots\>_G$ denotes the average over gauge configurations
(the trace, to be sure, is over spin and color).
From~\eqn{zeta_expl}  we read off
\bea
 [\zeta(\z)\psib(x)]_F    &=& U_0(0,\z) \,P^- \,G(z,x)|_{z_0=a_0} \, , \nn{}
{} [\psi(x)\bar{\zeta}(\y)]_F &=& G(x,y)|_{y_0=a_0} \,P^+ \,U_0(0,\y)^\dag \, .
\eea
To simplify~\eqn{ftwo} let us define
\beq 
 H(x) ~\equiv~ a^3 \sum_\y G(x,y)|_{y_0=a_0} \, P^+ \, U_0(0,\y)^\dag \, ,
\eeq
which is the solution of
\beq\label{Heqn}
  \sum_z Q(x,z) \cdot H(z) ~=~ {1\o a_0}\, \delta_{x_0,a_0} 
                 \, P^+ \, U_0(0,\x)^\dag \, , \qquad 0 < x_0 < T' \: ,
\eeq
and the boundary condition that $H(x)$ vanishes on the boundary layers.
Physically $H(x)$ is the propagator for a quark from anywhere on the
lower (outer) boundary to the point $x$ in the bulk. Note that
$H P^- \s= 0$, so only half of the components have to be calculated.

Using~\eqn{Gdag}  one finally obtains
\beq
 f_X(x_0) ~=~\half\, \<\, \Tr\, H^\dag(x)\, \ga_5\,\Gamma_X
                                           \, H(x) \, \>_G \: .
\eeq
Note that $\ga_5 \Gamma_X = 1$ for $X=P$, which means that $f_P(x_0)$
is positive configuration by configuration.

We used these formulas for the free case, solving~\eqn{Heqn} with
standard inversion algorithms,  to obtain the numbers
plotted in figures~\ref{fig:r} and~\ref{fig:th}.

As a check of some of our earlier remarks (and to see how small a volume
can be used in practice), we have repeated the determination of the
non-perturbative clover coefficient of the SW action on quenched 
$\beta\s= 6.0$ Wilson glue on a $12\cdot 6^3$ lattice at $aM=0.04$. We 
obtain $\om\s= 1.74(4)$, in agreement with the original calculation on 
a $16\cdot 8^3$ lattice at $aM \s= 0.01$~\cite{Sep}.

\section{Conclusions and Outlook}\label{sec:conclusion}

We are currently witnessing 
dramatic progress  in lattice gauge theory,
related to our increased understanding of the design of improved
actions.  More precisely, there are two important developments.
One is the ability to perform simulations
on coarse lattices of improved glue, the other the non-perturbative
elimination of $\Ord(a)$ errors in Wilson-type quark actions (and
operators), as suggested by the ALPHA collaboration.
Unfortunately, with quenched Wilson glue
the non-perturbative improvement of the SW action is not possible
on lattices coarser than $0.1$~fm; the quenched approximation seems to
break down for small quark masses on such lattices.
One thereby looses the great advantage of coarse lattices, the possibility
of performing a series of relatively cheap simulations on such lattices
that can then be accurately extrapolated to the continuum (the latter
assumes, of course, that $\Ord(a)$ errors are negligible).

The aim of this work has been to set the stage for the use of
non-perturbatively improved quark actions with improved glue,
by adapting the \SF{} to such actions. Much remains to be done.
We have sketched only one application, the tuning of the clover
coefficient(s), and even this area requires further study, certainly
on anisotropic lattices. 
One-loop calculations, although more difficult for improved gauge 
actions, should eventually be performed for various improvement 
coefficients.

The motivation for this work has been the hope that it is 
possible to non-perturbatively improve quark actions on significantly 
coarser lattices when using improved rather than Wilson glue.  
This indeed seems to be the case, as will
be described in~\cite{EHK} for the case of the SW action with
$a^2$ improved glue (in the quenched approximation).
Besides the SW action one should
also consider the D234 quark action, which has much smaller
violations of rotational symmetry. 
It will be interesting to see to what extent the occurrence of
``exceptional configurations'' is affected by the choice of quark
action. As mentioned in the introduction, our own experience 
with spectroscopic simulations indicates that the D234 action
has less near-zero modes (on the same configurations, at the same
quark mass).

Another option is the use of different kinds of improved glue. 
For the Wilson gauge action the breakdown of the quenched approximation
occurs on lattices that are still surprisingly fine (at least it seems 
so to us). A detailed understanding of this breakdown does not exist
yet, but it might, at least partially, be related to the (non-perturbative) 
phase structure of the Wilson action.
If such strong-coupling details of the action are relevant, certain types of 
improved glue 
might allow non-perturbative tuning on significantly
coarser lattices than others.

The problem of ``fake'' near-zero modes does
not exist in the continuum limit;      their occurrence is 
a lattice artifact related 
to the specific Wilson-type discretization used for the quark action
(the problem does not occur for staggered fermions). It would therefore
not really be surprising if improved actions have less problems with
exceptional configurations; such actions are, after all, supposed to be
more continuum-like. Note, for example, that adding higher derivatives
to an action makes the typical paths contributing to its path integral 
smoother
(cf.~e.g.~\cite{Simon}).\footnote{Improvement of a Wilson-type action
at $\Ord(a)$ is an exception in this respect; adding the clover
term leads to more, not less, problems with exceptional 
configurations~\cite{Sep}. It is only higher derivatives in the kinetic
term that lead to smoother typical paths.}
Of course, this statement is rather formal, and numerical studies of
these questions are called for.

A different strategy is to try to modify the quenched approximation
to eliminate the problem of exceptional configurations.
The challenge is to find a method that is systematic, practical, and
satisfactory on a conceptual level.
For a specific suggestion and some encouraging numerical studies, 
see~\cite{EC}. Such a procedure will most likely
allow non-perturbative tuning on much coarser lattices than possible
within the standard quenched approximation.

\vskip 1mm

We have discussed
the $\Ord(a)$ improvement of quark actions on anisotropic lattices, and
referred to the decisive advantages such lattices have for the study
of heavy quarks\footnote{Note that there are signs~\cite{Trot} that
NRQCD breaks down for charmonium. The D234 action on anisotropic
lattices would appear to be an ideal tool~\cite{ILQA,LAT96} for the study
of charmonium and heavy-light systems.}
and glueballs, for example. We suggested how to use the \SF{}
to achieve  ${\rm O}(a)$ improvement also on anisotropic
lattices. 

Finally, due to the great cost of full QCD simulations
we regard it as very important to study the improvement of quark actions 
coupled to {\it dynamical} improved glue, on isotropic and eventually also
anisotropic lattices.

We hope that the program outlined here will lead to the construction of
actions that allow accurate simulations of all aspects of QCD on coarse
lattices (this will also require the determination of various operator
renormalization constants along the lines pioneered by the ALPHA 
collaboration).
The advantages of coarse lattices make it imperative, we think,  to pursue 
these studies.

\vskip 12mm
\noindent
{\bf Acknowledgements}

\noindent    
I would like to thank Mark Alford, Khalil Bitar, Robert Edwards, 
Urs Heller, Tony Kennedy, Peter Lepage, and especially Stefan Sint 
for discussions. This work is supported by DOE grants 
DE-FG05-85ER25000 and DE-FG05-96ER40979.

\newpage

\appendix

\section{Notation and Conventions}\label{app:notation}

\subsection*{A.1~~~~In the Continuum}

We write the action of euclidean SU($N$) gauge theory in four dimensions as
\beq
 S_g[A] \, = \, {1\over 2 g^2} \int d^4x ~\Tr~ F_{\mu\nu}(x) F_{\mu\nu}(x) ~,
\eeq
where $F_{\mu\nu}(x)$ is the su($N$)-valued hermitean field strength. In our
conventions the covariant derivative is expressed in terms of the hermitean
gauge field $A_\mu(x)$ as  
\beq
 D_\mu ~=~ \p_\mu - i A_\mu
\eeq
so that
\beq
 F_{\mu\nu} ~=~ i [D_\mu,D_\nu] ~=~ \p_\mu A_\nu -\p_\nu A_\mu -i 
[A_\mu,A_\nu] ~.
\eeq
We use traceless hermitean su($N$) generators $T_a$, $a \s= 1,\ldots,N^2-1$,
normalized by $\Tr\, T_a T_b = {1\over 2}\de_{ab}$. We write
$F_{\mu\nu}(x) = F_{\mu\nu}^a(x) T_a$ and
$A_{\mu}(x) = A_{\mu}^a(x) T_a$.

The action of a Dirac fermion coupled to a gauge field is
\beq
 S_q[\psib,\psi,A] \= \int d^4x ~\psib(x) \,(D\slash + m) \,\psi(x) \,,
\eeq
where $D\slash = \sum_\mu \ga_\mu D_\mu$ in terms of the euclidean gamma 
matrices defined by
\bea
\ga_\mu &=& \ga_\mu^\ad \nn
\{ \ga_\mu, \ga_\nu \} &=& 2 \de_{\mu\nu} ~.
\eea
The spinor fields $\psi(x)$ and $\psib(x)$ also carry a suppressed color index
in the vector representation of SU($N$), and perhaps another suppressed
flavor index. 

The hermitean matrices $\si_\mn$ are defined by
$\si_{\mu\nu} = -{i\over 2}[ \ga_\mu,\ga_\nu ]$ or
$\ga_\mu \ga_\nu = \de_{\mu\nu} + i \si_{\mu\nu}$.
Note that
\beq\label{Dslsq}
 D\sl^2 \= \sum_\mu D_\mu^2 \+ \half \sum_\mn \si_\mn \Fmn \, .
\eeq
The term $\sigF \equiv \sum_{\mu\nu} \si_\mn \Fmn$ is known as the 
{\it clover} term (cf.~below).

\subsection*{A.2~~~~On the Lattice}

We will consider a  four-dimensional euclidean hypercubic lattice of extent
$L_\mu$ 
and lattice spacing $a_\mu$ in 
direction $\mu=0,1,2,3$.\footnote{Note that depending on how one
codes the \SF{} the number of time slices used in one's program, $L_0$,
will differ from the ``true physical extent'' $T$ of 
sect.~\ref{sec:gluons}.}
On a spatially symmetric lattice $L\equiv L_k$ for $k=1,2,3$.
For purely classical considerations the $a_\mu$ are arbitrary, but for
any situation where quantum effects play a role we will assume that the
lattice is spatially isotropic, $a \equiv a_k, \, k=1,2,3$.

Points are labelled by $x,y,z\ldots$,  as in the continuum. Spatial vectors
are denoted by boldface letters, as in $x\s= (x_0,\x)$.
We use $k,l$ for spatial indices.
 The notation $x\pm\mu$ is a shorthand
for $x\pm a_\mu \hat{\mu}$, where $\hat{\mu}$ is a unit vector in the positive
$\mu$-direction.

When working on anisotropic lattices one should in principle be   careful
in specifying the lattice spacing errors of various quantities. To avoid
cumbersome notation, we will
use $\Ord(a^n)$ to denote any combination of errors of the form
 $\Ord(a_\mu^n)$, up to logarithmic corrections. Only occasionally
do we use the notation  $\Ord(a_\mu^n)$ for an error depending only
on $a_\mu$, for some specific $\mu$.

The lattice gauge field, or link field, will be denoted by
$U_\mu(x)$ or $U_\mu(x_0,\x)$.      Mathematically it is a parallel
transporter from $x+\mu$ to  $x$, along a straight line (the link).
In terms of an underlying continuum gauge field $A_\mu(x)$ it can
be written as
\beq
 U_\mu(x) ~=~ \P \exp\biggl[ -i \int_x^{x+\mu} dx'_\mu A_\mu(x') \biggr] \, ,
\eeq
where $\P$ denotes path-ordering, stipulating that fields later
on the path are to be placed to the right of earlier fields.
We will employ  the notation $U_{-\mu}(x) \equiv U_\mu(x-\mu)^\dagger$ for the
parallel transporter from $x-\mu$ to $x$.

Our notation for lattice covariant derivatives was presented in sect.~3. 
We should add one detail here~\cite{May}. If one wants to impose ``twisted''
\BCs{} on the quark fields,
\beq
\psi( x+L_k \hat{k}) \= \e^{ i\theta_k} \psi(x) \, , \qquad
\psib(x+L_k \hat{k}) \= \e^{-i\theta_k} \psib(x) \, , 
\eeq
one can, equivalently, multiply each link field $U_k(x)$ in the derivative
operators of sect.~\ref{sec:quarks} by a phase $\exp(i a_k \theta_k/L_k)$,
while keeping the quark fields strictly periodic. 
This twist corresponds to adding an abelian gauge field to the action, so
that, for example, the covariant continuum derivative is given by
$D_\mu = \p_\mu - iA_\mu + i\theta_\mu/L_\mu$ ($\theta_0 = 0$).

When working up to $\Ord(a^3)$ errors in the quark action, classically, we can use
the {\it clover} representation of the lattice field strength
(see~e.g.~\cite{ILQA}) in the $\sigF$ term. This discretization differs at order 
$a^2$ from the continuum field strength.
To eliminate the $a^2$ errors in the field strength one can use an improved
discretization, for which we also refer to~\cite{ILQA}.

\section{Classical Boundary Errors of the Pure Gauge Schr\"odinger Functional}

We want to study the difference between the lattice 
and the continuum gauge actions with corresponding \BCs.
As remarked in sect.~\ref{sec:gluons}, this difference has two sources;
one being the difference between an integral and a sum, the other
the difference between the integrand and the summand.

When considering the continuum limit, classically or in perturbation
theory, one assumes the lattice field to be given by 
$U_\mu(x) = \P \exp[ -i \int_x^{x+\mu} dx'_\mu A_\mu(x')]$,
in terms of a {\it smooth} continuum field $A_\mu(x)$ with the appropriate
boundary conditions. Recall that the difference between the integral and
a sum over a smooth function with {\it periodic}
 boundary conditions is exponentially
suppressed in the lattice spacing; the exponential decay rate being equal
to the minimal distance of the singularities of the
integrand from the real axis (this follows from the Poisson summation
formula).

In the bulk, the lattice action densities differ from the continuum one at the
order $a^n$ of improvement. Any larger errors will be due the difference
between the sum and an integral at the boundary, or, possibly, due to a
difference between the continuum and lattice integrands at the boundary
(that is larger by more than a power of $a$ than the difference in the
bulk, since the effects of a boundary of codimension 1 are
suppressed by such a power). We will see that for improved actions both
errors contribute --- and in fact  partially 
cancel.

It will be useful to first describe the simpler case of the Wilson
action. In this case the difference between the lattice action density
and the continuum density $\propto \Tr\, F_\mn^2(x)$ is $\Ord(a^2)$ everywhere,
including the boundary, if we always associate the $\Tr\, F_\mn^2(x)$
arising from the expansion of a plaquette to the {\it middle} of the
plaquette.  For the temporal integral\footnote{Due
to the spatially periodic \BCs{} the space integral always comes along
for the ride and can be ignored.}
over the $F_{kl}^2$ terms arising from spatial plaquettes within
time slices, we then have a sum of the general form (cf.~figure~\ref{fig:wgts})
\beq\label{trapez}
 a \biggl[ \half f(0) + f(a) + \ldots + f(T\sm a) + \half f(T) \biggr ]
 \= \int_0^T dt \, f(t) \+ \Ord(a^2) 
\eeq
(to avoid cumbersome notation we use $a$ for the lattice spacing, even
though it should more properly be $a_0$ for a temporal integral).
This is the trapezoidal rule.

Similarly, for the $F_{0k}^2$ terms arising from temporal plaquettes
living between time slices, the midpoint rule tells us
\beq\label{midpoint}
 a \biggl[ f(a/2) + f(3a/2) + \ldots + f(T\sm 3a/2) + f(T\sm a/2) \biggr ]
 \= \int_0^T dt\, f(t) \+ \Ord(a^2)  \, .
\eeq
All in all we have shown that for the Wilson case there are classically
only $\Ord(a^2)$ boundary errors, if we choose the temporal weight factors as in 
figure~\ref{fig:wgts}. This was first shown in~\cite{LNWW} by a different
method.

Before proceeding with the improved case let us point out that the 
sum rules~\eqn{trapez} and~\eqn{midpoint} can be derived by 
``stringing together'' their elementary building blocks
\beq\label{trapez1}
 a \biggl[ \half f(0) + \half f(a) \biggr]
 \= \int_0^a dt\, f(t) \+ \Ord(a^3)
\eeq
and
\beq\label{midpoint1}
 a \, f(a/2)
 \= \int_0^a dt\, f(t) \+ \Ord(a^3) \, ,
\eeq
respectively.
These formulas are of course trivial; they and their slightly 
more sophisticated
cousins used below can all be derived in the same manner: Since the
integrand can (by assumption) be taylor expanded, one only has to prove
them for polynomials up to the appropriate order.

It will probably not have escaped the reader's attention that there is a
complete analogy between the elementary sum rules and their extended forms 
on the one hand, and the transfer matrix for one time step and  $T/a$ 
steps on the other.

Let us now apply the above lesson to the improved case. First of all
note that the continuum boundary layers should be placed in the middle
of the boundary double layers of figure~\ref{fig:wgts}. We will
relabel the time axis in this figure so that the continuum boundaries
are positioned at $t=0$ and $t=T$. The lattice time slices
now run from $t=-a_0/2$ to $t=T+a_0/2$ in steps of size $a_0$
(which will be denoted by $a$ again, from now on).

Consider first the errors associated with the spatial loops. Up to
$\Ord(a^4)$ errors we can associate an $F_{kl}^2$ term to the middle of
each spatial plaquette (by construction of the improved action). As in
the Wilson case, we can now proceed to analyze the sum over time slices
for fixed spatial $\x$. The question is whether there exists an
appropriate midpoint rule approximating an integral up to $\Ord(a^4)$
errors. The answer is positive,
\bea\label{midpoint_imp}
 && \hspace{-10mm} a \biggl[ 
 {1\o 24} f(-a/2)\sp {23\o 24} f(a/2)\sp f(3a/2) + \ldots 
 + f(T\sm 3a/2) \sp {23\o 24} f(T\sm a/2) \sp {1\o 24} f(T\sp a/2)\biggr ] \nn
 &=& \int_0^T dt\, f(t) \+ \Ord(a^4)  \,
\eea
which follows from the elementary rule
\beq\label{midpoint1_imp}
 a \biggl[ {1\o 24} f(-a/2) + {11\o 12} f(a/2) + {1\o 24} f(3a/2) \biggr ]
 \= \int_0^a dt\, f(t)  \+ \Ord(a^5) \, .
\eeq
Note that it is a priori quite surprising that it is possible to integrate
the polynomials $1, t, t^2, t^3$ exactly with only three sampling points.
In any case, we have shown that one should choose $b=1/24$ in 
figure~\ref{fig:wgts}.

We now come to the errors associated with temporal loops. These are a 
bit more subtle. The reason is the following: In the bulk the plaquettes
and rectangles combine to give, up to $\Ord(a^4)$ errors, an $F_{0k}^2$ term
situated at the middle of each plaquette; just as for the spatial loops.
 However, this not true for the
boundaries,\footnote{It were true if each of the tall-temporal rectangles 
that overlaps with the boundaries in figure~\ref{fig:wgts} were split
into two copies (each copy given a weight of $\half$), and one copy shifted
by one temporal lattice spacing so that it sticks 
out {\it beyond} the outer boundary layer of the lattice.}
which introduces an $a^2$ error. Note that this error is associated
with the {\it integrand} at the  boundary.
There is another $a^2$ boundary error, due to the
difference between the trapezoidal rule and the continuum integral.
It turns out that the corresponding correction term, exhibited in
the improved trapezoidal rule,
\beq\label{trapez_imp}
 a \biggl[ \half f(0) + f(a) + \ldots + f(T-a) + \half f(T) \biggr ]
 \- {a^2\o 12}(f'(T) - f'(0)) 
 \= \int_0^T dt\, f(t) + \Ord(a^4) 
\eeq
is exactly cancelled by the error from the integrand at the boundary!
We leave the explicit proof of this statement as an exercise to the reader
(who might want to consult~\cite{LWGlue} or~\cite{GPL}, for example,
for the small $a$ expansion of the plaquette and rectangle terms).

For completeness we remark that the improved trapezoidal rule follows
from the elementary rule
\beq\label{trapez1_imp}
 a \biggl[ 
 -{1\o 24} f(-a) + {13\o 24} f(0) + {13\o 24} f(a)-{1\o 24} f(2a) \biggr ]
 \= \int_0^a dt\, f(t) \+  \Ord(a^5)  
\eeq
(or from the Euler-Maclaurin formula).
This completes our proof that with $b=1/24$ in figure~\ref{fig:wgts}
the \SF{} for improved glue has no classical errors larger than $\Ord(a^4)$.

\section{Quark Actions on Anisotropic Lattices}

Consider an anisotropic lattice that is spatially isotropic.
Our main aim in this appendix is
to show that an on-shell $\Ord(a)$ improved Wilson-type quark
action on such a lattice can always be written as a discretization
of an effective continuum action of the form\footnote{A similar
statement, in a different but related context, can be found in~\cite{FNAL}.}
\beq\label{S_eff}
 S ~=~\int d^4x \, \psib(x) \Biggl[ m_0\+D\sl_0 \+ c \Dssl -\half r a_0 \,
 \biggl(\sum_\mu D_\mu^2 \+ \om_0 \sum_k \si_{0k} F_{0k} \+ \om 
\sum_{k<l} \si_{kl} F_{kl}\biggl) \Biggr ] \psi(x) \: .
\eeq
Here $D_\mu$ is the covariant (continuum) derivative, and $c$,
$\om_0$ and $\om$ are the parameters that have to be tuned at the
quantum level. $c$ is a bare velocity of light that has to be adjusted
so that the {\it renormalized} velocity of light is $1$, e.g.~by studying
the pion dispersion relation at small masses and momenta. $\om_0$ and
$\om$ are the temporal, respectively, spatial clover coefficients.
Classically all three coefficients are $1$, independent of the value
of the Wilson parameter $r$. On an isotropic lattice $c=1$, and only
$\om_0\equiv \om$ has to be  tuned.

To avoid doublers, the Wilson parameter can in principle be chosen
to have any value $r>0$ (for the $W_1$ case 
there is an upper bound of $\Ord(1)$ from
requiring reflection positivity; the exact value depends on the 
specific discretization used). Conceptual and 
practical considerations usually lead
to a specific choice of $r$ for any given discretization;
cf.~\eqn{br_values}. The non-perturbative values of $\om_0$, $\om$
and $c$ depend on $r$, of course. 

Note that to $\Ord(a)$  we can without loss of generality always assume that 
$\om_0$ and $\om$ are mass-independent. 
On the other hand, $c$ can have
 a mass dependence on the quantum level at $\Ord(a)$.

Before proving the above claims, it will be useful to provide a bit
of background on {\it classically} improved, doubler-free quark
actions on anisotropic lattices. The simplest way to construct such
actions~\cite{ILQA} is to start with a continuum action
$\psib_c \, Q_c \, \psi_c \equiv \psib_c (D\slash + m_c)\psi_c$ and perform
the field redefinition
$\psi_c = \Om_c \, \psi, ~\psib_c =\psib \, \Omb_c$ with
\beq\label{cvstd}
\Omb_c \= \Om_c ~,  \quad 
 \Omb_c \, \Om_c \= 1 \- {r a_0 \over 2} \, (D\slash - m_c) ~.
\eeq
The subscripts `c' stand for `continuum'. A field redefinition is just
a change of variable in the path integral, so on-shell quantities
are not affected.\footnote{The
Jacobian of a field transformation matters only at the quantum 
level, where, in the case at hand, its leading effect is to
renormalize the gauge coupling.}
The transformed quark matrix reads
\bea\label{QOm}
 \Omb_c \, Q_c \, \Om_c ~=~ m_c (1 + \half r a_0 m_c) \+ D\slash
       - {1\over 2} \, r a_0 \,
   \Bigl(\sum_\mu D_\mu^2 \+ \half \sigF \Bigr) ~,
\eea
where we used eq.~\eqn{Dslsq}.

Note that the above still defines a continuum action;  $a_0$ should
at this point just be considered a formal dimensionful parameter.
It takes on the meaning of a temporal lattice spacing only in the
next step: We obtain the W$_1$ and W$_2$ actions of sect.~\ref{sec:quarks} 
simply by replacing the continuum operators
$D\slash, a_0 D_\mu^2$ and $a_0 \Fmn$ by suitable lattice
versions, differing at $\Ord(a^n)$  from the former, with
$n\s= 2$ for W$_1$ and $n\s= 3$ for W$_2$. For a more detailed
exposition see~\cite{ILQA}.

By construction, classical lattice actions obtained in this manner have 
on-shell errors only at $\Ord(a^n)$.\footnote{Note, in particular, that if
one writes $m_0 = m_c(1+\half r a m_c)$ in eq.~\eqn{Wactions}, then
$m_c$ agrees with the pole mass up to $a^n$ errors.}
 The proves our claim relating  to~\eqn{S_eff} in the classical case. 

Note that $\Ord(a^n)$ errors can also
be achieved for off-shell quantities simply by {\it undoing} the 
field redefinition (on the lattice now), to obtain continuum-like 
lattice fields. By abuse of notation we will also denote the lattice 
fields for the action obtained by discretization of~\eqn{QOm} as
$\psi$, $\psib$ (like the fields defined by
the inital change of variable in the continuum).
For  quark bilinears, where things are particularly simple, we can then
rewrite bilinears of the lattice fields in terms of continuum fields as
\bea\label{bilinear}
 \psib(x) \, \Gamma \, \psi(x) &=& 
 \psib_c(x) \,\Omb_c^{-1} \,\Gamma \,\Om_c^{-1} \,\psi_c(x) \+ \Ord(a^n) \nn[0.7mm]
 &=& (1+r a_0 m_c)^{-1} \, \psib_c(x) \,\Gamma \, \psi_c(x) \+ \Ord(a^n) \: .
\eea
In the above we used the field equation for $\psi_c$  to eliminate
$D\sl$ in the field transformation operators. 
We see that for bilinears classical improvement simply amounts to the 
multiplication by a mass-dependent factor (that, furthermore, is the same 
for all bilinears).

Let us now return to our claim for the quantum case at $\Ord(a)$.
The possible terms that can appear at this order in the effective action
are of the form 
\beq
D_0^2, \quad \sum_k D_k^2, \quad [\Dssl,D\sl_0],  \quad 
\sum_k \si_{0k} F_{0k},  \quad 
    \sum_{k<l} \si_{kl} F_{kl} \: .
\eeq
Remember also that at $\Ord(a^0)$ the
relative coefficient of the temporal and spatial derivatives can
renormalize. 

To prove our claim it suffices to show that starting with
the continuum action $\int \psib [m_c +D\sl_0 + c' \Dssl ]\psi$ we can
find a field redefinition that generates the $D_0^2$, $D_k^2$ and
$[\Dssl,D\sl_0]$ terms in~\eqn{S_eff} with {\it arbitrary} coefficients;
then we can always adjust the values of these coefficients to equal
those in~\eqn{S_eff} (e.g.~zero for the  $[\Dssl,D\sl_0]$ term).
In other words, $D_0^2$, $D_k^2$ and $[\Dssl,D\sl_0]$ are {\it redundant}
operators.

We assume that we have used our freedom of rescaling the fields to
ensure that the coefficient of $D\sl_0$ is $1$ at $\Ord(a^0)$.
The most general field redefinition at $\Ord(a)$ is then
\bea\label{cvgen}
 \Om  &=& 1 \- {r_0 a_0\o 4} D\sl_0 \- {r a\o 4} \Dssl \+{r_m a_0 \o 4} m_c \nn
 \Omb &=& 1 \- {\bar{r}_0 a_0\o 4} D\sl_0 \- {\bar{r} a\o 4} \Dssl 
              \+ {\bar{r}_m a_0 \o 4} m_c \: .
\eea
This leads to
\bea
&& \Omb \, (m_c \+ D\sl_0 \+ c' \Dssl ) \, \Om \nn
&&= m_c \, [1+{a_0\o 4} m_c (r_m+\bar{r}_m)] \, \+ \,
         D\sl_0 \, [1+{a_0\o 4} m_c (r_m+\bar{r}_m-r_0-\bar{r}_0)] \nn
&&\quad \+ \Dssl \, [c' + {a\o 4} m_c (c'(r_m+\bar{r}_m)-r-\bar{r})] 
     \,\- \,  {a_0\o 4}(r_0+\bar{r}_0) \, D_0^2 
     \,\- \,  {a\o 4}(r+\bar{r})\,  c' \,\Dssl^2 \nn
&&\quad \- {1\o 8} \,
            [c' a_0 (r_0+\bar{r}_0) + a(r+\bar{r})] \, \{\Dssl, D\sl_0\}\nn
&&\quad \- {1\o 8} \, 
            [c' a_0 (r_0-\bar{r}_0) - a(r-\bar{r})] \, [\Dssl, D\sl_0] \:.
\eea
Remembering $\Dssl^2 = \sum_k D_k^2 + \sum_{k<l} \si_{kl} F_{kl}$ and
$\{\Dssl,D\sl_0\} = \sum_k \si_{0k} F_{0k}$, we see that
our claim follows if we parameterize the independent free parameters of the
change of variable as $r_0+\bar{r}_0$ (to adjust the $D_0^2$ coefficient),
$r+\bar{r}$ (to adjust $\sum_k D_k^2$),
$r-\bar{r}$ (to adjust $ [\Dssl, D\sl_0]$) and
$r_m+\bar{r}_m$ (to eliminate any possible mass dependence
of the $D\sl_0$ coefficient at $\Ord(a)$). 
No further simplifications are possible by suitable
choice of $r_0-\bar{r}_0$ or $r_m-\bar{r}_m$; the latter does not even
appear in the above. 

The mass dependence of $c$ at $\Ord(a)$ is a complication in practice.
To minimize systematic uncertainties it is preferable to always
determine the clover coefficients $\om_0$ and $\om$  at zero
quark mass. To do so one needs to know $c$ at zero quark mass. It
remains to be seen if the most naive method of determining this value ---
demanding the pion dispersion relation to be relativistic at small momenta,
 and then extrapolating to zero pion mass --- is sufficiently accurate.
One expects that $c$ has only a weak dependence on $\om_0$ and $\om$;
otherwise, one might have to repeat the above procedure iteratively
until self-consistent values of $c$, $\om_0$ and $\om$ are obtained.

\end{document}